\documentclass{article}

\newif\ifmstarpreprint
\mstarpreprintfalse

\PassOptionsToPackage{numbers, compress}{natbib}
\ifmstarpreprint
  \usepackage[preprint]{neurips_2026}
\else
  \usepackage[main,final]{neurips_2026}
\fi
\usepackage[utf8]{inputenc}
\usepackage[T1]{fontenc}
\usepackage{CJKutf8}

\usepackage{hyperref}
\hypersetup{hidelinks}
\usepackage{url}
\usepackage{booktabs}
\usepackage{tabularx}
\usepackage{amsfonts}
\usepackage{amssymb}
\usepackage{amsmath}
\usepackage{nicefrac}
\usepackage{microtype}
\usepackage[table]{xcolor}
\usepackage{algorithm}
\usepackage{algorithmic}
\usepackage{graphicx}
\usepackage{caption}
\usepackage{wrapfig}
\usepackage{placeins}
\usepackage{listings}
\usepackage{pifont}
\usepackage{fontawesome5}
\usepackage{tikz}
\usetikzlibrary{calc,shapes.geometric}
\ifmstarpreprint
  \usepackage[most]{tcolorbox}
  \usepackage{varwidth}
\fi
\usepackage{enumitem}
\setlist{leftmargin=*}

\ifmstarpreprint
\definecolor{abstractbg}{gray}{0.97}
\renewenvironment{abstract}{%
  \begin{tcolorbox}[
    enhanced,
    breakable,
    colback=abstractbg,
    colframe=black!15,
    boxrule=0.4pt,
    arc=4pt,
    left=6pt,
    right=6pt,
    bottom=6pt,
    top=6pt,
    shadow={1pt}{-1pt}{0pt}{black!8},
  ]
  \small\noindent\tcbox[on line, colback=black!8, colframe=black!8, boxrule=0pt, arc=2pt, left=2pt, right=2pt, top=1pt, bottom=1pt]{\textbf{\normalsize Abstract}}\enspace
}{%
  \end{tcolorbox}
}
\fi

\definecolor{codebg}{HTML}{F7FAFC}
\definecolor{codeframe}{HTML}{93C5FD}
\definecolor{codeaccent}{HTML}{2563EB}
\definecolor{codecomment}{HTML}{64748B}
\definecolor{codestring}{HTML}{047857}
\definecolor{codekeyword}{HTML}{1D4ED8}

\lstdefinestyle{evolvedcode}{
  language=Python,
  basicstyle=\ttfamily\scriptsize\linespread{1.05}\selectfont,
  keywordstyle=\bfseries\color{codekeyword},
  commentstyle=\itshape\color{codecomment},
  stringstyle=\color{codestring},
  numbers=none,
  frame=single,
  frameround=tttt,
  framesep=5pt,
  framerule=0.45pt,
  rulecolor=\color{codeframe},
  backgroundcolor=\color{codebg},
  breaklines=true,
  showstringspaces=false,
  columns=flexible,
  keepspaces=true,
  aboveskip=0.75em,
  belowskip=0.6em,
  xleftmargin=0.6em,
  xrightmargin=0.6em,
  framexleftmargin=0.55em,
  framexrightmargin=0.55em,
  framextopmargin=0.25em,
  framexbottommargin=0.25em,
  captionpos=t,
  escapeinside={(*@}{@*)},
  morekeywords={self},
}

\newcommand{\sysname}{\textsc{Mstar}}
\newcommand{\second}[1]{\textcolor{green!50!black}{\textbf{#1}}}

\title{M\textsuperscript{$\bigstar$}: Every Task Deserves Its Own Memory Harness}

\author{
  \textbf{Wenbo Pan}$^{1}$ \quad
  \textbf{Shujie Liu}$^{2}$ \quad
  \textbf{Xiangyang Zhou}$^{2}$ \quad
  \textbf{Xianlong Wang}$^{1}$\thanks{Correspondence to: \texttt{xianlong.wang@my.cityu.edu.hk}} \\[0.3em]
  \textbf{Shiwei Zhang}$^{2}$ \quad
  \textbf{Bingjing Xu}$^{2}$ \quad
  \textbf{Wanlu Shi}$^{2}$ \quad
  \textbf{Xiaohua Jia}$^{1}$ \\[0.5em]
  $^{1}$City University of Hong Kong \qquad
  $^{2}$Microsoft
}

\begin{document}

\maketitle

\begin{center}
\vspace{-3em}
\includegraphics[width=0.86\textwidth]{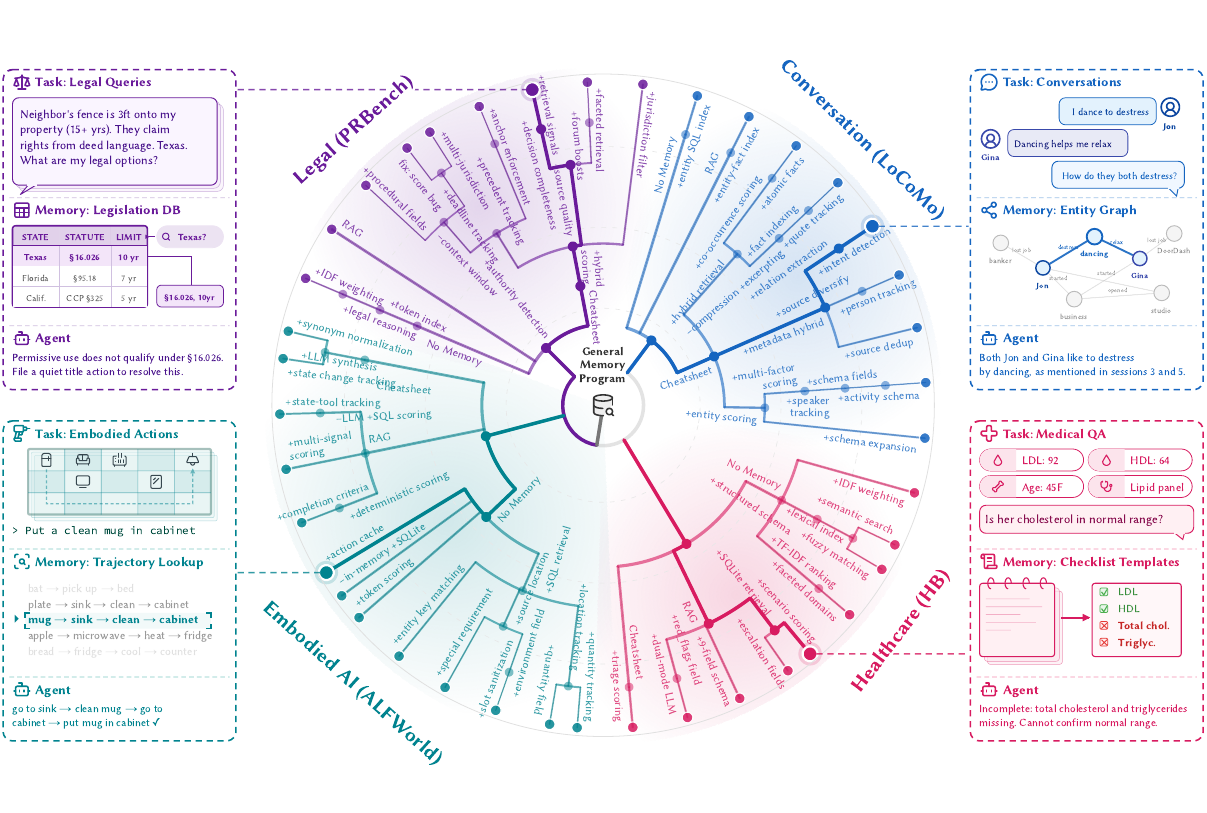}
\captionsetup{type=figure,hypcap=false}
\caption{\textbf{Evolved memory harnesses across tasks.}
The radial tree shows how \sysname{} evolves shared seed programs into task-specialized memory harnesses for LoCoMo, ALFWorld, HealthBench, and PRBench.
The surrounding panels show representative task, memory, and agent behavior for each domain.}
\label{fig:system-overview}
\end{center}
\vspace{-1em}

\begin{abstract}
Large Language Model (LLM) agents rely on a memory harness to write, organize, retrieve, and use past experience.
A harness that works for one task often fails on another, because different tasks like conversations, embodied planning, or expert reasoning require different storage and retrieval behavior.
To address this limitation, we introduce \sysname{}, a method that automatically discovers task-optimized memory programs through reflective code evolution.
Specifically, \sysname{} represents a memory harness as an executable Python program that defines the \textit{Schema}, \textit{Logic}, and \textit{Instruction} components, which are optimized jointly.
We evaluate \sysname{} on four tasks covering conversation, embodied planning, and specialized reasoning.
Our results demonstrate that \sysname{} improves performance over existing static memory harnesses across all evaluated tasks.
Furthermore, the evolved memory programs exhibit structurally distinct processing mechanisms for each evaluated domain.
These findings suggest that every task benefits from its own memory harness, and that memory-program search provides a concrete way to discover it.
\par\medskip
\noindent\textbf{Code:} \href{https://github.com/wbopan/mstar}{\textcolor{codeaccent}{\texttt{github.com/wbopan/mstar}}} \quad\textbf{Live demo:} \href{https://mstar.wenbo.io}{\textcolor{codeaccent}{\texttt{mstar.wenbo.io}}}\label{absend}
\end{abstract}

\section{Introduction}

Agentic systems powered by large language models (LLMs) need a memory harness to reuse past experience across diverse tasks~\citep{zhang2024survey_memory,park2023generative,zhong2024memorybank,wang2024augmenting}.
We use \textit{memory harness} to refer to the components that store, organize, retrieve, and use this experience.
As different task domains require knowledge in different structures, they call for different memory harnesses.
For example, conversational agents need semantic maps of people, events, and preferences to remember user-specific facts (Figure~\ref{fig:system-overview}, top right)~\citep{locomo}, while embodied AI agents need reusable patterns from past trajectories to choose actions in new environments (Figure~\ref{fig:system-overview}, bottom left)~\citep{alfworld}, and specialized reasoning domains may need structured databases (Figure~\ref{fig:system-overview}, top left)~\citep{prbench}.
These requirements encode different assumptions about what to store, how to organize it, and when to retrieve it.
A harness built around one set of assumptions can therefore transfer poorly to another~\citep{structmemeval}.
This raises a practical question: \textit{How can we find the most suitable memory harness for a given task?}

To answer this question, we propose \sysname{}, an evolutionary method that automatically discovers task-optimized memory harnesses for LLM agents.
Specifically, \sysname{} represents a memory harness as a \textit{memory program}: a Python module with a fixed interface consisting of three components, \textit{Schema}, \textit{Logic}, and \textit{Instruction}.
The \textit{Schema} defines the data formats for storage and retrieval, the \textit{Logic} implements the \texttt{read} and \texttt{write} operations, and the \textit{Instruction} guides the agent's interaction with the knowledge base.
These components are jointly optimized through reflective code evolution.
Prior works have also improved agent memory harnesses from experience~\citep{zhao2024expel,cai2025flex,memskill,gepa}.
Most closely, MemEvolve~\citep{memevolve} first showed that treating the memory system itself as an evolvable program can significantly improve its performance on downstream tasks.
\sysname{} builds on this direction but has a different emphasis.
Rather than improving one memory system that works across tasks, it jointly optimizes Schema, Logic, and Instruction to find a specialized harness for each target task.
To search this space, \sysname{} uses two mechanisms: a reflector agent that debugs memory program failures and generates targeted improvements, and a population-based program search strategy that balances exploration and exploitation while enforcing constraints on program quality.

We apply \sysname{} to four tasks covering conversation, embodied planning, and specialized reasoning, evaluating its performance against ten competitive baselines across four memory paradigms~\citep{locomo,alfworld,healthbench,prbench}.
Our results indicate that \sysname{} achieves the best overall score and performs robustly on every task; in contrast, the baselines perform well only on specific tasks.
Furthermore, we observe that the optimal memory programs for different tasks are distinct in both structure and procedure.
By exploring a broad memory-program search space, \sysname{} evaluates diverse memory programs to identify the task-optimized memory program.
Together, these results support the view that every task benefits from its own memory harness, discovered as an executable memory program rather than chosen as a fixed design.

In summary, our main contributions are as follows:
\begin{enumerate}
    \item We formulate task-optimized memory harness discovery as \textit{memory-program search}, where an executable Python program jointly specifies Schema, Logic, and Instruction.
    \item We propose \sysname{}, an evolutionary method that optimizes memory harnesses from target-task feedback through reflective code evolution and achieves the strongest overall performance across four tasks compared to competitive generalist memory harnesses.
    \item We perform in-depth analysis over the memory harness space and find that task-optimized memory harnesses are distinct in both structure and procedure across different tasks: conversation, embodied planning, and expert reasoning tasks all require different optimal memory harnesses.
\end{enumerate}

\begin{figure*}[t]
\centering
\includegraphics[width=\textwidth]{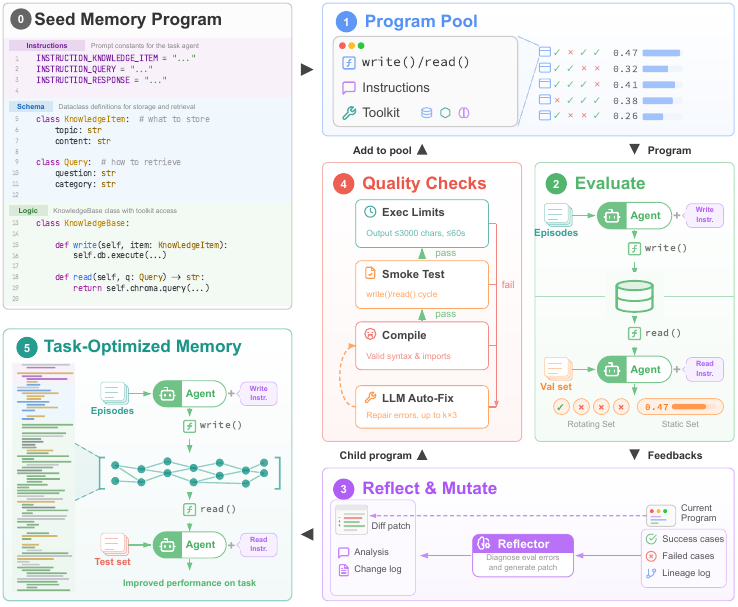}
\caption{\textbf{System overview of \sysname{}.}
Numbered stages show the workflow from seed memory program~(\textbf{0}) to final test evaluation~(\textbf{5}): population pool~(\textbf{1}), task evaluation~(\textbf{2}), reflective code evolution~(\textbf{3}), and compile/runtime quality checks~(\textbf{4}).}
\label{fig:method-overview}
\vspace{-1em}
\end{figure*}

\section{Problem Setting}
\label{sec:problem-setting}

We consider a task $\mathcal{T} = (\mathcal{D}_\text{e}, \mathcal{D}_\text{v}, \mathcal{D}_\text{t})$, where $\mathcal{D}_\text{e} = \{e_1, \dots, e_N\}$ is a set of episode trajectories representing past experiences, $\mathcal{D}_\text{v} = \{(q_1, y_1), \dots, (q_M, y_M)\}$ is a validation set of queries with corresponding ground-truth answers, and $\mathcal{D}_\text{t}$ is a held-out test set used only for final reporting.
An agent $\mathcal{A}$ must memorize information from these past episodes to answer queries from $\mathcal{D}_\text{v}$ and $\mathcal{D}_\text{t}$.
Specifically, the agent is equipped with a knowledge base $\mathcal{K}$; it observes episodes in $\mathcal{D}_\text{e}$ and extracts knowledge items $k$ to populate this knowledge base $\mathcal{K}$.
At test time, given a query $q_j$, the agent accesses $\mathcal{K}$, which returns relevant context $c_j = \mathcal{K}.\texttt{read}(q_j)$.
Building on this retrieved information, the agent then produces its answer or takes an action conditioned on both the query and the context:
\begin{equation}
\hat{y}_j = \mathcal{A}(q_j, c_j).
\label{eq:inference}
\end{equation}

\paragraph{Problem Definition: Task-Optimized Memory Program.}
While several prior works have proposed fixed memory methods for agent systems, these generalist structures transfer poorly across task types~\citep{structmemeval}.
To this end, we formalize task-optimized memory harness optimization as search over executable memory programs.
Each program $P$ defines the input and output formats of the knowledge base, its underlying organization method such as a vector database or relational tables, and the agent's policy for when and how to access memory.
Consequently, $P$ describes an open program space that can express many storage and retrieval procedures.
We define $\mathcal{J}(P)$ as the agent's performance on $\mathcal{D}_\text{v}$ when equipped with a knowledge base instantiated by $P$.
The objective is to find the optimal memory program $P^*$ that maximizes this performance metric:
\begin{equation}
P^* = \arg\max_P \mathcal{J}(P).
\label{eq:objective}
\end{equation}
In the next section, we address how we can automatically discover the optimal memory program for a given task. Specifically, we propose a framework that uses reflective code evolution to search this program space by iterating and making targeted improvements starting from a seed program.

\section{Method}
\label{sec:mstar}

Figure~\ref{fig:method-overview} gives an overview of the system \sysname{}, a reflective code evolution method that automatically discovers task-optimized memory programs.
Specifically, \sysname{} consists of three main components: (1) representing the memory-program search space in code (Section~\ref{sec:search-space}), (2) reflective code evolution (Section~\ref{sec:reflection}), and (3) population-based program search (Section~\ref{sec:population}).

\subsection{Representing the Memory-Program Search Space in Code}
\label{sec:search-space}

We represent each memory program as an executable Python module with a fixed interface.
To define a broad memory-program search space, each memory program simultaneously controls three key components of the knowledge base: \textit{Schema}, \textit{Logic}, and \textit{Instruction}.
In addition, the program has access to a standardized \textit{Toolkit}. A template for a memory program is shown in Figure~\ref{fig:method-overview}.

\begin{itemize}
    \item \textbf{\textit{Schema}.}\looseness=-1
    The Schema specifies the data formats that the knowledge base accepts for writing and querying.
    For example, the input format can range from structured data with timestamps to natural language insights.
    We implement the Schema using Python dataclasses.
    As a result, when the agent needs to write or query information, it instantiates these dataclass types and passes them into the memory program for parsing.

    \item \textbf{\textit{Logic}.}
    The Logic defines the backend operations of the knowledge base when processing and storing Schema-formatted inputs and queries.
    This component can include operations such as computing embeddings, managing SQL tables, or calling an LLM for secondary data processing.
    Specifically, the Logic exposes \texttt{write} and \texttt{read} functions, which are executed whenever the agent performs write or read actions.

    \item \textbf{\textit{Instruction}.}
    The Instruction defines how the agent interacts with the knowledge base.
    This includes how to extract information from observed episodes, how to formulate queries, and how to interpret the results returned by the knowledge base.
    We define the Instruction through constant prompt strings.
    These instruction constants are inserted into the system prompt when the agent executes corresponding write, query, and response actions.

    \item \textbf{Toolkit.}
    Within the memory-program search space, we allow the memory program to use various data structures and external tools.
    We provide a set of whitelisted components, including lists, heaps, relational databases, vector databases, and LLM endpoints.
    The Toolkit enables memory programs to implement complex storage and retrieval procedures.
\end{itemize}

\subsection{Reflective Code Evolution}
\label{sec:reflection}

Representing a memory program using executable code places memory-program search in an enormous search space.
Inspired by recent evolutionary algorithm discovery methods, we design a reflective code evolution process that iteratively refines memory programs using validation feedback~\citep{romera2024funsearch,chen2023evoprompting,deepmind2025alphaevolve}.
Specifically, this process consists of three stages: sampling feedback from a validation loop, iterative refinement via a coding agent, and constraint checking with automated repair before the candidate is evaluated. An illustration is shown in Figure~\ref{fig:method-overview}.

\textbf{Sampling feedback from the validation loop.}
We use a set of episodes and validation queries to evaluate a memory program.
The validation queries are partitioned into a \textit{static} set, which remains constant across all iterations, and a \textit{rotating} set, which changes in each iteration similar to mini-batches in training.
During validation, the agent initiates a query to the knowledge base for each sample in both the rotating and static sets and uses the returned information to generate a response or take an action.
The aggregated score on the static validation set serves as the performance metric for the current memory program.
Meanwhile, the generated trajectories on the rotating set are used as feedback to improve the program later.

\textbf{Coding agent iteration.}
We employ a coding agent to iteratively update the current memory program to improve performance and fix potential bugs.
We provide the coding agent the execution trajectories from the read and write processes, underperforming samples from the rotating validation set, and the change logs from prior iterations along with their corresponding scores.
Based on this information, the coding agent analyzes whether the read and write operations function as expected, identifies the root causes of underperforming cases, and determines which past improvements were effective.
Consequently, the coding agent generates a code patch to apply to the current program and produces a new change log to guide future iterations.
The complete prompt template is provided in Appendix~\ref{app:prompts}.

\textbf{Constraint checks and automated repair.}
Before applying the updated memory program, the code evolution process executes a set of runtime checks to ensure its quality.
If any check fails, the memory program and its corresponding error messages are resubmitted to the coding agent for repair. These checks prevent low-quality programs from consuming excessive computational resources, and any program that still fails after a fixed number of repair attempts is discarded.

\subsection{Population-Based Program Search}
\label{sec:population}

Unlike function search or prompt optimization, evaluating a memory program in \sysname{} can be highly expensive.
For any program change, the agent must regenerate episode knowledge and evaluate the program on validation samples.
Therefore, the search procedure should explore the program space efficiently under a limited evaluation budget.

Instead of continuously updating a single program, which can cause the evolution algorithm to fall into local optima, we maintain a program pool.
We initialize the search with a small set of simple seed programs (Appendix~\ref{app:seeds}).
In each iteration, we sample a parent program from the pool with probability biased toward higher validation scores.
Specifically, the probability $\Pr(P_i)$ of selecting program $P_i$ from the pool is computed as:
\begin{equation}
    \Pr(P_i) = \frac{\exp(\mathcal{J}(P_i) / \tau)}{\sum_{j} \exp(\mathcal{J}(P_j) / \tau)},
\end{equation}
where $\mathcal{J}(\cdot)$ is the static validation score (Eq.~\ref{eq:objective}) and $\tau$ is the softmax temperature.
The coding agent then applies a reflective mutation step to the selected parent, and the resulting child program is added back to the pool.
This design balances exploration and exploitation, as top-performing programs have a greater chance of being selected for improvement, yet lower-scoring programs retain a non-zero probability of being explored.
We use additional subset-selection procedures to reduce evaluation cost, and describe these implementation details in Appendix~\ref{app:representative-subset}.

Overall, \sysname{} provides an efficient reflective code evolution procedure for memory-program search.
In the following section, we provide an empirical analysis to validate our method.

\begin{table*}[t]
    \centering
    \footnotesize
    \setlength{\tabcolsep}{3.5pt}
    \caption{\textbf{Main results across four task families.}
    Columns report LoCoMo token F1 and LLM-judge score (L-J), ALFWorld success rate, HealthBench rubric score, and PRBench rubric score.
    Baselines are grouped by memory paradigm.
    Test sets contain 100 items for LoCoMo and each HealthBench split and 50 items for each ALFWorld and PRBench split.
    Best per column in \textbf{bold}, second best in \second{green}.
    $^\dagger$ marks a significant lead in a paired bootstrap test ($p<0.05$, Appendix~\ref{app:significance}).
    L-J is not evaluated for MemEvolve.}
    \vspace{-0.5em}
    \label{tab:main-results}
    \newcolumntype{Y}{>{\centering\arraybackslash}X}
    \begin{tabularx}{\textwidth}{l *{2}{Y} *{2}{Y} *{2}{Y} *{2}{Y}}
        \toprule
        \rowcolor{gray!10}
         & \multicolumn{2}{c}{\textbf{LoCoMo}} & \multicolumn{2}{c}{\textbf{ALFWorld}} & \multicolumn{2}{c}{\textbf{HealthBench}} & \multicolumn{2}{c}{\textbf{PRBench}} \\
        \cmidrule(lr){2-3} \cmidrule(lr){4-5} \cmidrule(lr){6-7} \cmidrule(lr){8-9}
        \rowcolor{gray!10}
        \textbf{Method} & \textbf{F1} & \textbf{L-J} & \textbf{Unseen} & \textbf{Seen} & \textbf{Data} & \textbf{Emerg.} & \textbf{Legal} & \textbf{Finance} \\
        \midrule
        No Memory            & 0.036 & 0.030 & 0.738 & 0.640 & 0.242 & 0.429 & 0.431 & 0.269 \\
        \midrule
        \multicolumn{9}{l}{\textit{\textbf{Retrieval-based Systems} (Memorizing Episodes)}} \\
        \midrule
        \quad Vector Search   & 0.256 & 0.400 & 0.643 & 0.720 & 0.264 & 0.400 & 0.466 & 0.308 \\
        \quad G-Memory        & 0.224 & 0.380 & 0.690 & 0.560 & 0.309 & 0.447 & 0.499 & 0.318 \\
        \quad Mem0            & \second{0.373} & \second{0.540} & 0.738 & 0.640 & 0.216 & 0.413 & 0.450 & 0.321 \\
        \midrule
        \multicolumn{9}{l}{\textit{\textbf{Self-evolution Systems} (Memorizing Experiences)}} \\
        \midrule
        \quad Trajectory Retrieval & 0.276 & 0.420 & 0.714 & \second{0.780} & 0.265 & 0.442 & 0.480 & 0.304 \\
        \quad ReasoningBank        & 0.194 & 0.380 & 0.738 & 0.580 & 0.315 & 0.470 & 0.508 & 0.330 \\
        \quad Dynamic Cheatsheet   & 0.124 & 0.190 & 0.619 & 0.480 & 0.286 & \second{0.487} & 0.474 & 0.327 \\
        \midrule
        \multicolumn{9}{l}{\textit{\textbf{Prompt-optimizing Systems} (Learned Prompts)}} \\
        \midrule
        \quad GEPA                 & 0.132 & 0.190 & \second{0.857} & 0.720 & 0.304 & 0.466 & \second{0.568} & \second{0.449} \\
        \quad GEPA + Vector Search & 0.300 & 0.390 & 0.810 & \textbf{0.820} & \second{0.327} & 0.484 & 0.554 & 0.411 \\
        \midrule
        \multicolumn{9}{l}{\textit{\textbf{Program-evolving Systems} (Evolved Memory Code)}} \\
        \midrule
        \quad MemEvolve            & 0.204 & -- & 0.760 & 0.620 & 0.267 & 0.456 & 0.484 & 0.317 \\
        \midrule
        \rowcolor{green!10}
        \textbf{\sysname{}} & \textbf{0.459}$^\dagger$ & \textbf{0.610} & \textbf{0.881}$^\dagger$ & 0.700 & \textbf{0.390}$^\dagger$ & \textbf{0.493} & \textbf{0.660}$^\dagger$ & \textbf{0.586}$^\dagger$ \\
        \bottomrule
    \end{tabularx}%
    \vspace{-1em}
\end{table*}

\section{Experimental Setup}
\label{sec:experimental-setup}

\looseness=-1
We evaluate \sysname{} on four benchmark families with seven domain configurations, covering long-term conversational question answering, embodied task completion, medical question answering, and professional legal and financial reasoning.
For each configuration, we run 20 evolution iterations, select the best memory program on validation, and report performance on a held-out test split.
We use GPT-5.4 Mini as the task agent and GPT-5.3-Codex as the coding agent in reflective code evolution~\citep{openai2026gpt54mini,openai2026gpt53codex}.
Additional dataset splits, hyperparameters, evaluation protocols, computational costs, and pseudocode are provided in Appendix~\ref{app:datasets}, Appendix~\ref{app:hyperparameters}, Appendix~\ref{app:eval-protocols}, Appendix~\ref{app:cost}, and Appendix~\ref{app:algorithm}.

\noindent\textbf{Benchmarks.}
We use LoCoMo~\citep{locomo} to test long-term conversational memory, ALFWorld~\citep{alfworld} to test memory use in embodied planning, HealthBench~\citep{healthbench} to test rubric-graded health reasoning, and PRBench~\citep{prbench} to test professional legal and financial reasoning.
Together, these benchmarks cover retrieval-heavy question answering, interactive decision making, and specialized reasoning under structured evaluation.
Dataset construction details are provided in Appendix~\ref{app:datasets}, and metric details are provided in Appendix~\ref{app:eval-protocols} for all seven benchmark configurations.

\noindent\textbf{Baselines.}
We compare \sysname{} with ten baselines, including a no-memory control and nine memory-based methods spanning four paradigms, as summarized in Table~\ref{tab:main-results}.
These baselines cover direct answering without external memory, retrieval-based memory over stored observations, self-evolving memories that distill reusable experience, prompt-optimization methods with and without vector retrieval, and MemEvolve, which also evolves memory systems as code~\citep{rag,alma,ouyang2025reasoningbank,gepa,memevolve}.
Except for Vector Search and No Memory, which serve as generic references, all baselines are published, human-designed memory systems or optimizers.
All evolutionary baselines use the same task agent, coding model, data splits, and number of evolution rounds as \sysname{}.
Detailed baseline implementations are provided in Appendix~\ref{app:baselines}, with implementation notes for every baseline.

\begin{figure*}[t]
\centering
\includegraphics[width=\textwidth]{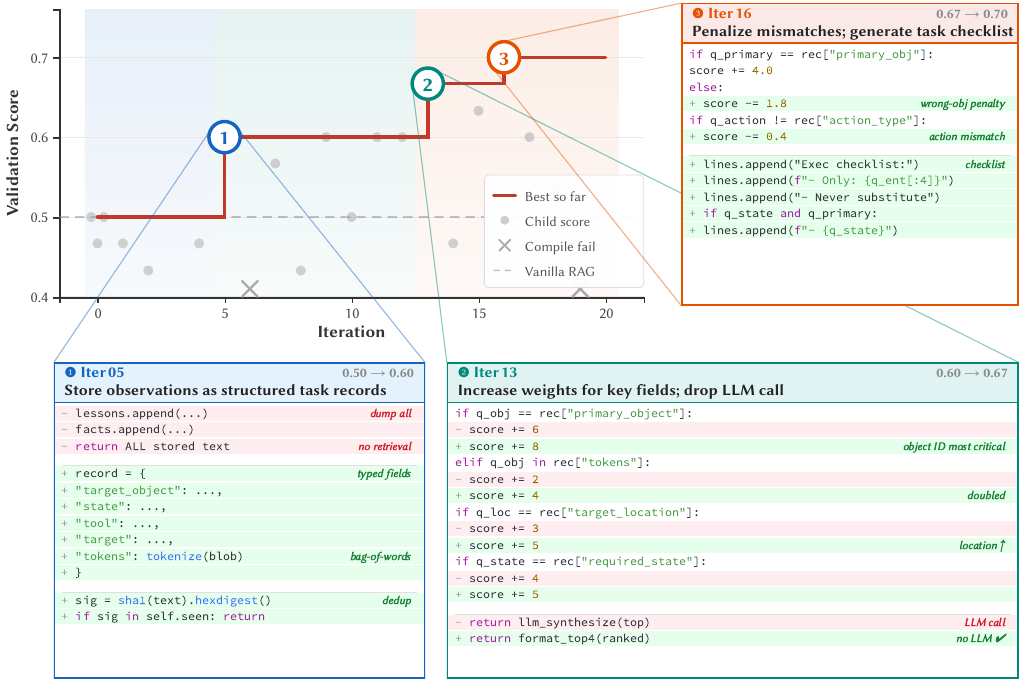}
\caption{\textbf{ALFWorld evolution trajectory.}
The main panel plots each child program's validation score and the best score so far across 20 evolution iterations, with crosses marking compile failures and the dashed line marking the Vanilla RAG baseline.
Numbered panels highlight representative code changes that structure task records, reweight retrieval, and add mismatch penalties and execution checklists for selecting the next action.}
\label{fig:evolution-trajectory}
\vspace{-1em}
\end{figure*}

\section{Results}
\label{sec:results}

We organize the analysis around three questions: whether memory-program search improves task performance, whether the resulting programs differ structurally across tasks, and which evolved components drive the gains.

\paragraph{Reflective code evolution progressively discovers better memory programs.}\looseness=-1
Table~\ref{tab:main-results} shows that \sysname{} achieves the highest score on seven of eight reported metrics.
Compared with the best baseline in each winning metric, the relative improvement is up to 31\%, with the largest gains on LoCoMo and PRBench.
The strongest baseline varies by task, indicating that existing memory methods have task-dependent strengths.
GEPA is competitive on ALFWorld and PRBench, while Mem0, GEPA+Vector Search, and Dynamic Cheatsheet lead the baselines on LoCoMo and HealthBench.
In comparison, \sysname{} achieves the best overall performance.
MemEvolve, which also evolves memory code with the same coding model, scores below \sysname{} on all seven configurations and is closest on HealthBench and PRBench.
On five of the six configurations where \sysname{} leads, its lead is statistically significant (Appendix~\ref{app:significance}).
Appendix~\ref{app:additional} further shows that the gain comes mainly from search rather than from the coding model's prior, that \sysname{} reaches the final validation score of GEPA with a fraction of its budget, and that the lead holds under an independent judge and other task models.
This pattern suggests that executable code search reaches useful memory designs beyond the strongest fixed baselines.
Figure~\ref{fig:evolution-trajectory} shows the ALFWorld evolution trajectory.
The evolved memory program improves steadily over iterations, with the reflector producing task-specific changes at different stages of search.
Early edits build a usable task-record schema that captures objects, locations, and state changes from demonstrations.
Later edits refine retrieval weights and add action-selection checks, allowing the agent to retrieve more relevant trajectory patterns and choose actions more reliably.

\paragraph{Different tasks produce structurally distinct optimal memory programs.}
\begin{wraptable}{r}{0.44\textwidth}
    \vspace{-1em}
    \centering
    \footnotesize
    \caption{\textbf{Ablation study on LoCoMo.}
    Rows report the full system and variants that remove or alter one design choice; metric is token F1.
    $\Delta$ reports absolute change from the full system.}
    \label{tab:ablation}
    \begin{tabularx}{\linewidth}{@{}>{\raggedright\arraybackslash}Xcc@{}}
        \toprule
        \rowcolor{gray!10}
        \textbf{Variant} & \textbf{F1} & \textbf{$\Delta$} \\
        \midrule
        \rowcolor{green!10}
        \sysname{}                           & 0.459  & ---      \\
        \quad \textit{$-$\,Instruction}       & 0.353  & $-$0.106 \\
        \quad \textit{$-$\,Code}              & 0.256  & $-$0.203 \\
        \quad \textit{Max parent selection}   & 0.381  & $-$0.078 \\
        \quad \textit{$-$\,Diversity}         & 0.318  & $-$0.141 \\
        \bottomrule
    \end{tabularx}%
    \vspace{-1em}
\end{wraptable}

\begin{figure*}[t]
\centering
\includegraphics[width=\textwidth]{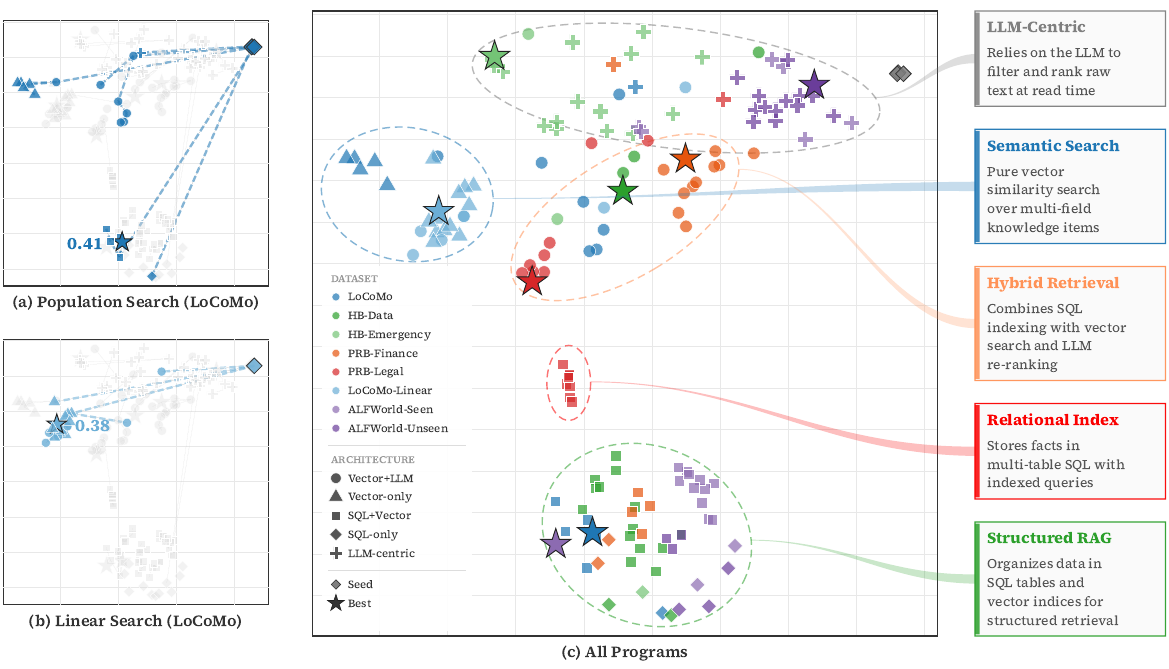}
\caption{%
  \textbf{Program embedding landscape.}
  Each evolved program is embedded with a code embedding model and projected to 2D via t-SNE~\citep{vandermaaten2008tsne}.
  \textbf{(a,\,b)}~Population-based search (\sysname{}) and linear search, with colored edges tracing parent-child lineage.
  \textbf{(c)}~All programs from the evaluated configurations and the LoCoMo linear-search ablation, colored by configuration.
  Marker shapes denote the storage architecture discovered by evolution: circles for vector\,+\,LLM, triangles for vector-only, squares for SQL\,+\,vector, rotated squares for SQL-only, and crosses for LLM-centric designs; diamonds mark seed programs and stars mark the best program for each configuration.
}
\label{fig:landscape}
\end{figure*}

To understand how \sysname{} adapts memory program structure to different tasks, we visualize the program embedding landscape accumulated during evolution (Figure~\ref{fig:landscape}).
For each task and iteration, we first sanitize the program by replacing strings and variable names with placeholders so that the embedding emphasizes structural differences.
We then embed the sanitized programs with a code embedding model and project them with t-SNE~\citep{vandermaaten2008tsne}.
Because positions in this projection depend on the embedding model and t-SNE, we use the figure only to group programs by task and storage architecture, not to trace how programs evolve.
The projection places the best programs for different benchmark configurations in separate regions, with markers indicating different storage architectures.
For example, the best ALFWorld program uses a compact SQLite action cache with read-time LLM synthesis.
The best LoCoMo program uses ChromaDB semantic retrieval together with SQLite lexical and metadata scoring.
Appendix~\ref{app:evolved-comparison} reports structural summaries, and the supplementary material includes full code for all best programs.

\begin{figure*}[t]
\centering
\includegraphics[width=\textwidth]{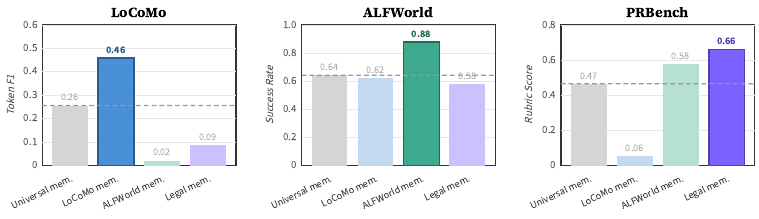}
\caption{\textbf{Cross-task transfer of evolved memory programs.}
Each panel fixes one target benchmark and compares memory programs evolved on different source benchmarks.
Highlighted bars mark native-task programs, and the dashed line marks the universal seed baseline.}
\label{fig:task-optimized-memory}
\vspace{-1em}
\end{figure*}

To further test whether these structures are task-specific, we evaluate the best evolved program from each source task on other target tasks (Figure~\ref{fig:task-optimized-memory}).
In all three targets, the native evolved program outperforms transferred programs from other benchmarks.
Five of six transferred programs perform worse than the universal seed baseline.
This pattern suggests that memory program structure should be optimized together with the target task.

\paragraph{Joint evolution of structure and policy enables task-specific adaptation.}
Table~\ref{tab:ablation} isolates key design choices on LoCoMo.
Each variant removes one component of \sysname{}: \textit{$-$\,Instruction} lets only the code evolve and keeps the prompt instructions fixed; \textit{$-$\,Code} does the opposite, evolving only the prompt instructions while keeping the code at the Vector Search seed; \textit{Max parent selection} always picks the highest-scoring parent instead of sampling by softmax; and \textit{$-$\,Diversity} drops the population pool and mutates only the current best program at each iteration.
\textit{$-$\,Code} gives the largest drop, reducing F1 by 0.203.
\textit{$-$\,Instruction} reduces F1 by 0.106.
The full system also scores above \textit{Max parent selection} and \textit{$-$\,Diversity}.
These results indicate that joint optimization of program Logic and Instruction gives the strongest LoCoMo result.

\paragraph{Evolution is stable across random seeds.}

\begin{wraptable}{r}{0.64\textwidth}
    \vspace{-1.2em}
    \centering
    \scriptsize
    \caption{\textbf{Stability across evolution seeds.}
    Mean and std.\ are computed over five independent stability runs with random seeds; metrics match the primary columns in \autoref{tab:main-results}.
    CV is the coefficient of variation in percent, Best Baseline is the strongest fixed baseline, and Win is the number of seeds above that baseline.}
    \vspace{-0.5em}
    \label{tab:stability}
    \begin{tabularx}{\linewidth}{@{}l >{\centering\arraybackslash}X >{\centering\arraybackslash}p{0.6cm} >{\centering\arraybackslash}p{2.6cm} >{\centering\arraybackslash}p{0.6cm}@{}}
        \toprule
        \rowcolor{gray!10}
        \textbf{Benchmark} & \textbf{Mean{\scriptsize$\pm$Std}} & \textbf{CV} & \textbf{Best Baseline} & \textbf{Win} \\
        \midrule
        LoCoMo          & $0.421_{\pm0.042}$ & 10.0 & 0.373 {\scriptsize(Mem0)}         & 4/5 \\
        HB Data         & $0.391_{\pm0.023}$ &  5.9 & 0.327 {\scriptsize(GEPA{+}VS)}    & 5/5 \\
        PR Finance      & $0.561_{\pm0.032}$ &  5.7 & 0.449 {\scriptsize(GEPA)}          & 5/5 \\
        \bottomrule
    \end{tabularx}%
    \vspace{-1em}
\end{wraptable}

\looseness=-1
A practical concern for evolutionary methods is sensitivity to random seeds, especially under a 20-iteration budget in which discovery quality can depend on early mutations.
To evaluate this effect, Table~\ref{tab:stability} reports test scores from five independent runs on three benchmarks.
Across all benchmarks, the coefficient of variation remains at or below 10\%, which indicates stable outcomes with moderate variance.
Across the 15 runs, 14 outperform the strongest baseline, supporting consistent gains across initializations.
Appendix~\ref{app:stability} provides full per-seed scores and validation trajectories.

\section{Discussion}
\label{sec:discussion}

In this section, we examine two properties of reflective code evolution: search diversity and category effects.

\paragraph{How does evolution explore the program space?}

Figure~\ref{fig:landscape}(c) provides another view of the search process.
Evolved programs form several regions with different storage designs, including LLM-centric, vector-only, SQL-only, SQL+vector, and vector+LLM programs.
Programs from different benchmarks often appear in the same regions.
This pattern suggests that evolution can try common storage designs before adapting them to a task.
Figure~\ref{fig:landscape}(a,b) further compares population-based search with a linear variant.
The linear variant starts from one empty seed and repeatedly mutates only the current best program, which corresponds to the \textit{$-$\,Diversity} ablation in Table~\ref{tab:ablation}.
Population search keeps multiple candidates active, so later mutations can draw from more than one candidate lineage.
The LoCoMo ablation matches this interpretation: \textit{$-$\,Diversity} reduces test F1 from 0.459 to 0.318, and \textit{Max parent selection} reaches 0.381 (Table~\ref{tab:ablation}).
Together, these results indicate that the memory-program space is diverse and multi-modal.
Population-based search explores this space effectively by maintaining multiple candidate lineages, improving the chance of finding a strong task-specific program under a small iteration budget.

\paragraph{Does evolution improve performance uniformly?}

\begin{table*}[t]
    \centering
    \small
    \caption{\textbf{Per-category performance breakdown.}
    \textbf{(a)}~ALFWorld Unseen success rate by task type.
    \textbf{(b)}~LoCoMo token F1 by query type.
    $n$ is the number of test episodes or test queries per category.
    Min reports the minimum score across shown categories.
    Best per column in \textbf{bold}.}
    \vspace{-0.5em}
    \label{tab:per-category}
    \resizebox{\textwidth}{!}{%
    \begin{tabular}{@{}l ccccc c !{\vrule width 0.6pt} cccc c@{}}
        \toprule
        \rowcolor{gray!10}
        & \multicolumn{6}{c!{\vrule width 0.6pt}}{\textbf{(a) ALFWorld Unseen}} & \multicolumn{5}{c}{\textbf{(b) LoCoMo}} \\
        \cmidrule(r){2-7} \cmidrule(l){8-12}
        \rowcolor{gray!10}
        & \textbf{Simple} & \textbf{Clean} & \textbf{Cool} & \textbf{Heat} & \textbf{2-Obj} &
        & \textbf{Single} & \textbf{Temp.} & \textbf{Causal} & \textbf{Open} & \\
        \rowcolor{gray!10}
        & \footnotesize$(n{=}8)$ & \footnotesize$(n{=}11)$ & \footnotesize$(n{=}7)$ & \footnotesize$(n{=}6)$ & \footnotesize$(n{=}8)$ & \textbf{Min}
        & \footnotesize$(n{=}18)$ & \footnotesize$(n{=}18)$ & \footnotesize$(n{=}9)$ & \footnotesize$(n{=}55)$ & \textbf{Min} \\
        \midrule
        No Memory            & 0.88 & 0.82 & 0.57 & 0.83 & 0.62 & 0.57 & 0.02 & 0.00 & 0.11 & 0.04 & 0.00 \\
        Mem0                 & 0.88 & 0.64 & 0.57 & \textbf{1.00} & 0.75 & 0.57 & \textbf{0.33} & 0.26 & 0.37 & 0.43 & 0.26 \\
        Trajectory Retrieval & 0.75 & 0.73 & 0.71 & 0.67 & 0.75 & 0.67 & 0.18 & 0.33 & 0.34 & 0.28 & 0.18 \\
        GEPA + Vector Search & 0.88 & \textbf{1.00} & 0.57 & \textbf{1.00} & 0.62 & 0.57 & 0.15 & \textbf{0.38} & \textbf{0.39} & 0.31 & 0.15 \\
        \rowcolor{green!10}
        \sysname{}      & 0.88 & 0.91 & \textbf{1.00} & 0.83 & 0.75 & \textbf{0.75} & \textbf{0.33} & 0.19 & 0.35 & \textbf{0.51} & 0.19 \\
        \bottomrule
    \end{tabular}}%
    \vspace{-1em}
\end{table*}

Table~\ref{tab:per-category} disaggregates the results in Table~\ref{tab:main-results} by category (full per-category tables are in Appendix~\ref{app:per-category}).
On ALFWorld Unseen, \sysname{} raises the worst category score to 0.75, above the best baseline floor among the methods shown (0.67).
This pattern suggests that the evolved action cache addresses failure modes that fixed designs leave exposed across task types.
On LoCoMo, the gains are less uniform: \sysname{} is strongest on open-domain questions and tied on single-hop recall, while Mem0 and GEPA+Vector Search remain stronger on temporal and causal questions.
This split suggests that an aggregate objective can improve the overall score while leaving lower-frequency or harder categories underserved.

\section{Related Work}

\noindent\textbf{Memory for LLM Agents.}
Prior work on agent memory builds external stores from interaction histories~\citep{zhang2024survey_memory}: typical pipelines extract salient information, retrieve relevant entries for a new query, and update via consolidation or pruning~\citep{park2023generative,zhong2024memorybank,wang2024augmenting}.
Recent systems learn richer policies; MemSkill~\citep{memskill} selects natural-language skills via an RL controller, and AWM~\citep{wang2024awm} induces reusable workflows from successful trajectories.
These systems commit to a fixed memory representation at design time.
MemEvolve~\citep{memevolve} is the closest prior work to ours.
It treats the memory system as an evolvable program, decomposes it into encode, store, retrieve, and manage modules, and lets a coding agent rewrite these modules, so its search space also covers a broad class of memory designs.
\sysname{} shares this view and also evolves executable memory code with a coding agent and constraint checks.
However, MemEvolve focuses on memory designs that generalize across tasks and evaluates them on deep-research benchmarks, while \sysname{} searches for a separate harness for each target task and evaluates it on conversation, embodied planning, and expert reasoning tasks.

\noindent\textbf{Self-Evolving Agents.}~\looseness=-1
Self-evolving LLM agents improve from interaction experience without gradient-based updates~\citep{gao2025survey_selfevolving}.
ExpeL~\citep{zhao2024expel} distills trajectories into editable natural-language insights, FLEX~\citep{cai2025flex} builds an experience library via continual reflection, and a complementary line automates agent design through meta-optimization (ADAS~\citep{hu2024adas}), skill synthesis (Voyager~\citep{wang2023voyager}), and verbal self-critique (Reflexion~\citep{shinn2023reflexion}).
These systems evolve text-level artifacts that the LLM must re-interpret at runtime; \sysname{} instead evolves deterministic code, reducing interpretation variance and enabling explicit data structures and retrieval logic in each memory program.

\noindent\textbf{Program Search with Language Models. }
Using LLMs as mutation operators has yielded solutions surpassing human heuristics in combinatorial optimization~\citep{romera2024funsearch}, neural architecture search~\citep{chen2023evoprompting}, and algorithmic discovery~\citep{deepmind2025alphaevolve}.
Among prompt-level methods, GEPA~\citep{gepa} mutates every prompt in a compound AI system via reflective natural-language updates with Pareto-front candidate selection.
\sysname{} applies the same paradigm to executable memory programs; because generated code can fail to compile or violate runtime constraints, we add a compile-fix loop and runtime violation recovery that are unnecessary in prompt evolution (Section~\ref{sec:reflection}).

\section{Limitations and Future Work}
\label{sec:limitations}

\noindent\textbf{Limitations.}
\sysname{} has several limitations.
First, the search signal can reward behavior that raises the metric without producing a strictly better harness.
For example, the LLM judges of HealthBench and PRBench tend to prefer long answers, and the evolved harnesses on these benchmarks also instruct the agent to give long and detailed answers.
Similarly, although the reflector prompt discourages hard-coded word lists, evolution still proposes string literals copied from the dataset text on every benchmark, most often on PRBench-Legal (Appendix~\ref{app:hardcoding-audit}).
Such literals can improve in-task performance, but they also limit transfer to other tasks (Figure~\ref{fig:task-optimized-memory}).
Second, \sysname{} does not help on every configuration.
On ALFWorld-Seen it trails GEPA + Vector Search, although its 95\% interval contains the score of GEPA + Vector Search (Appendix~\ref{app:significance}).
On LoCoMo, systems designed specifically for long-term conversation, such as LightMem~\citep{lightmem} and SimpleMem~\citep{simplemem}, report higher scores than the evolved harness, while our baselines were chosen to cover different memory paradigms rather than to be the strongest LoCoMo systems.
Third, all evolution runs use one task agent and reflector pair (GPT-5.4 Mini and GPT-5.3-Codex).
The evolved harnesses keep their lead under two other task models (Appendix~\ref{app:cross-model}), but we have not evolved harnesses with a different or open-source coding model.
Finally, each run re-ingests episodes and re-scores the validation set for every candidate, so one 20-iteration run costs up to \$90 on rubric-graded benchmarks and up to 100 wall-clock hours on ALFWorld (Appendix~\ref{app:cost}).

\noindent\textbf{Future work.}
Because a previous harness can serve as the seed program, a natural next step is to warm-start evolution from a harness evolved on a related task and measure how much this reduces search cost.
We also plan to stress-test evolved harnesses with noisy, stale, or contradictory memories, and to study search procedures and objectives that prefer reusable structure over benchmark-specific heuristics.

\section{Conclusion}

We introduced \sysname{}, a method that discovers task-optimized memory programs for large language model agents through reflective code evolution.
By representing a memory harness as executable Python code with Schema, Logic, and Instruction, \sysname{} turns memory design into a memory-program search problem.
Across four benchmarks, the evolved programs outperform strong memory baselines and develop distinct storage, retrieval, and action workflows for different domains.
These results support the view that agent memory should be optimized for the target task, and that memory-program search provides a practical path toward this goal.
\begin{ack}
This work was supported by HK RGC Research Grants No.\ RFS2425-1S01 and C1043-24GF.
\end{ack}
\label{endmain}

\bibliographystyle{plainnat}
\bibliography{references}

\appendix

\section*{Appendix}

\vspace{0.4em}

\newcommand{\appsection}[3]{%
  \noindent\colorbox{gray!7}{\parbox{\dimexpr\linewidth-2\fboxsep}{%
    \textbf{\textcolor{cyan!70!black}{#1}\hspace{0.5em}#2}\hfill\pageref{#3}%
  }}\par\vspace{0.05em}%
}
\newcommand{\appsubsection}[2]{%
  \noindent\hspace{1em}#1\dotfill\pageref{#2}\par%
}

\begingroup
\scriptsize
\begin{center}
\textbf{\large Table of Contents}
\end{center}

\vspace{0.25em}
\hrule
\vspace{0.35em}

\appsection{A}{Reflective Code Evolution Algorithm}{app:algorithm}
\appsubsection{A.1\hspace{0.5em}Evolution Loop}{alg:evolution-loop}
\appsubsection{A.2\hspace{0.5em}Split Evaluation}{alg:split-eval}

\vspace{0.1em}
\appsection{B}{Benchmarks and Dataset Details}{app:datasets}
\appsubsection{B.1\hspace{0.5em}Dataset Splits}{app:dataset-splits}
\appsubsection{B.2\hspace{0.5em}LoCoMo}{app:dataset-locomo}
\appsubsection{B.3\hspace{0.5em}ALFWorld}{app:dataset-alfworld}
\appsubsection{B.4\hspace{0.5em}HealthBench}{app:dataset-healthbench}
\appsubsection{B.5\hspace{0.5em}PRBench}{app:dataset-prbench}

\vspace{0.1em}
\appsection{C}{Baseline Methods}{app:baselines}
\appsubsection{C.1\hspace{0.5em}No Memory}{app:baseline-no-memory}
\appsubsection{C.2\hspace{0.5em}Retrieval-Based Systems}{app:baseline-retrieval}
\appsubsection{C.3\hspace{0.5em}Self-Evolution Systems}{app:baseline-self-evolution}
\appsubsection{C.4\hspace{0.5em}Prompt-Optimizing Systems}{app:baseline-prompt}
\appsubsection{C.5\hspace{0.5em}Program-Evolving Systems}{app:baseline-program}

\vspace{0.1em}
\appsection{D}{Experiment Configuration}{app:experiment-config}
\appsubsection{D.1\hspace{0.5em}Computational Cost}{app:cost}
\appsubsection{D.2\hspace{0.5em}Hyperparameters}{app:hyperparameters}
\appsubsection{D.3\hspace{0.5em}Representative Subset Selection}{app:representative-subset}

\vspace{0.1em}
\appsection{E}{Seed Programs}{app:seeds}
\appsubsection{E.1\hspace{0.5em}Vector Search}{app:seed-vector-search}
\appsubsection{E.2\hspace{0.5em}LLM Summarizer}{app:seed-llm-summarizer}
\appsubsection{E.3\hspace{0.5em}Experience Learner}{app:seed-experience-learner}

\vspace{0.1em}
\appsection{F}{Evolved Programs}{app:evolved-programs}
\appsubsection{F.1\hspace{0.5em}ALFWorld: Deterministic Action Cache}{app:evolved-alfworld}
\appsubsection{F.2\hspace{0.5em}LoCoMo: Multi-Signal Episodic Index}{app:evolved-locomo}
\appsubsection{F.3\hspace{0.5em}HealthBench and PRBench Program Sketches}{app:evolved-health-pr}
\appsubsection{F.4\hspace{0.5em}Cross-Benchmark Structural Comparison}{app:evolved-comparison}

\vspace{0.1em}
\appsection{G}{Prompt Templates and Mutation Interface}{app:prompts}
\appsubsection{G.1\hspace{0.5em}Task Agent Prompts}{app:task-agent-prompts}
\appsubsection{G.2\hspace{0.5em}Reflector Prompt}{app:reflector-prompt}
\appsubsection{G.3\hspace{0.5em}Compile-Fix Prompt}{app:compile-fix-prompt}

\vspace{0.1em}
\appsection{H}{Evaluation Protocols}{app:eval-protocols}
\appsubsection{H.1\hspace{0.5em}Token F1 (LoCoMo)}{app:metric-token-f1}
\appsubsection{H.2\hspace{0.5em}Binary Success (ALFWorld)}{app:metric-binary-success}
\appsubsection{H.3\hspace{0.5em}Rubric-Based Scoring}{app:metric-rubric}

\vspace{0.1em}
\appsection{I}{Extended Results}{app:extended-results}
\appsubsection{I.1\hspace{0.5em}Complete Per-Category Results}{app:per-category}
\appsubsection{I.2\hspace{0.5em}Multi-Seed Stability Results}{app:stability}

\vspace{0.1em}
\appsection{J}{Additional Analyses}{app:additional}
\appsubsection{J.1\hspace{0.5em}Significance of the Main Results}{app:significance}
\appsubsection{J.2\hspace{0.5em}Search versus the Coding Model's Prior}{app:single-shot}
\appsubsection{J.3\hspace{0.5em}Search Efficiency}{app:search-efficiency}
\appsubsection{J.4\hspace{0.5em}Transfer across Task Models}{app:cross-model}
\appsubsection{J.5\hspace{0.5em}Independent Judge}{app:judge}
\appsubsection{J.6\hspace{0.5em}Transfer to LongMemEval}{app:longmemeval}
\appsubsection{J.7\hspace{0.5em}Audit of Hard-Coded Literals and Evaluation Loopholes}{app:hardcoding-audit}
\appsubsection{J.8\hspace{0.5em}Overlap between Splits}{app:split-overlap}

\vspace{0.1em}
\appsection{K}{Broader Impact}{app:broader-impact}

\vspace{0.1em}
\appsection{L}{LLM Usage}{app:llm-usage}

\vspace{0.3em}
\hrule
\endgroup

\section{Reflective Code Evolution Algorithm}
\label{app:algorithm}

\begin{algorithm}[!htbp]
\caption{\textbf{Reflective code evolution for memory-program search.}
The left panel gives the population-based evolution loop, and the right panel gives split evaluation for a candidate program.}
\label{alg:evolution}
\begin{minipage}[t]{0.47\linewidth}
\textbf{(a) Evolution Loop} \label{alg:evolution-loop}
\begin{algorithmic}[1]
\REQUIRE Seed programs $\{S_1,\ldots,S_K\}$, budget $B$, temp.\ $\tau$
\REQUIRE Episodes $D_\text{e}$, static val $D_\text{v}$, rotate pool $D_\text{r}$
\ENSURE Best program $P^*$
\STATE $\mathcal{P} \gets \emptyset$
\FOR{$i = 1$ \TO $K$}
    \STATE $(J_i, \_) \gets \textsc{Eval}(S_i, D_\text{e}, D_\text{v})$
    \STATE $\mathcal{P} \gets \mathcal{P} \cup \{(S_i, J_i)\}$
\ENDFOR
\FOR{$t = 1$ \TO $B$}
    \STATE $P_\text{par} \gets \textsc{SoftmaxSample}(\mathcal{P}, \tau)$
    \STATE $\tilde{D} \gets \textsc{Sample}(D_\text{r},\, n_r)$
    \STATE $(\_, \mathcal{R}) \gets \textsc{Eval}(P_\text{par}, D_\text{e}, \tilde{D})$
    \STATE $P_\text{mut} \gets \textsc{Mutate}(P_\text{par}, \mathcal{R})$
    \STATE $P' \gets \textsc{CompileFix}^{k}(P_\text{mut})$ \COMMENT{$\bot$ if discarded}
    \IF{$P' \neq \bot$}
        \STATE $(J', \_) \gets \textsc{Eval}(P', D_\text{e}, D_\text{v})$
        \STATE $\mathcal{P} \gets \mathcal{P} \cup \{(P', J')\}$
    \ENDIF
\ENDFOR
\RETURN $\arg\max_{(P,J) \in \mathcal{P}} J$
\end{algorithmic}
\end{minipage}%
\hfill
\begin{minipage}[t]{0.50\linewidth}
\textbf{(b) Split Evaluation} \label{alg:split-eval}
\begin{algorithmic}[1]
\REQUIRE Program $P$, task agent $\mathcal{A}$, metric $\mu$, threshold $\theta$
\REQUIRE Episodes $D_\text{e}$, eval queries $D_\text{eval}$
\ENSURE Score $J$, diagnostic cases $\mathcal{R}=(\mathcal{F},\mathcal{S})$
\STATE \textit{// Phase 1: Knowledge ingestion}
\STATE $\text{KB} \gets P.\textsc{KnowledgeBase}(\text{toolkit})$
\FOR{$d \in D_\text{e}$}
    \STATE $\text{item} \gets \mathcal{A}.\texttt{extract}(d,\; P.\textsc{InstructionKnowledgeItem})$
    \STATE $\text{KB}.\texttt{write}(\text{item},\, d)$
\ENDFOR
\STATE \textit{// Phase 2: Retrieval and scoring}
\STATE $\mathcal{F}, \mathcal{S} \gets \emptyset, \emptyset$
\FOR{$(q, y) \in D_\text{eval}$}
    \STATE $\text{query} \gets \mathcal{A}.\texttt{formulate}(q,\; P.\textsc{InstructionQuery})$
    \STATE $c \gets \text{KB}.\texttt{read}(\text{query})$
    \STATE $\hat{y} \gets \mathcal{A}.\texttt{respond}(q, c,\; P.\textsc{InstructionResponse})$
    \STATE $s \gets \mu(\hat{y},\, y)$
    \IF{$s < \theta$}
        \STATE $\mathcal{F} \gets \mathcal{F} \cup \{(q, y, \hat{y}, c, s)\}$
    \ELSE
        \STATE $\mathcal{S} \gets \mathcal{S} \cup \{(q, y, \hat{y}, c, s)\}$
    \ENDIF
\ENDFOR
\RETURN $\bigl(\tfrac{1}{|D_\text{eval}|}\sum s,\;\; (\mathcal{F}, \mathcal{S})\bigr)$
\end{algorithmic}
\end{minipage}
\end{algorithm}

\FloatBarrier

\section{Benchmarks and Dataset Details}
\label{app:datasets}

\subsection{Dataset Splits}
\label{app:dataset-splits}

\autoref{tab:datasets} summarizes the data splits for each benchmark configuration.

\begin{table}[!htbp]
\centering
\caption{\textbf{Dataset split sizes.}
Rows report train, validation, and test sizes for every benchmark split used by the evaluation pipeline.}
\label{tab:datasets}
\small
\begin{tabular}{@{}llrrr@{}}
\toprule
\textbf{Benchmark} & \textbf{Split} & \textbf{Train} & \textbf{Val} & \textbf{Test} \\
\midrule
LoCoMo         & ---               & 272 sessions  & 1{,}440 QA & 100 QA \\
\midrule
ALFWorld       & Unseen            & 4{,}652 trajectories & 145 ep.    & 50 ep. \\
               & Seen              & 4{,}652 trajectories & 150 ep.    & 50 ep. \\
\midrule
HealthBench    & Data Tasks        & 257           & 120        & 100 \\
               & Emergency         & 260           & 122        & 100 \\
\midrule
PRBench        & Legal             & 100           & 100        & 50 \\
               & Finance           & 120           & 130        & 50 \\
\bottomrule
\end{tabular}
\end{table}

For LoCoMo, we exclude category~5 (adversarial and unanswerable questions), which tests refusal capability rather than memory retrieval, retaining 1{,}540 QA pairs across four categories~\citep{locomo}.
We reserve 100 QA pairs for the held-out test split and use the remaining 1{,}440 QA pairs as validation data.
For PRBench, hard items (\texttt{finance\_hard} and \texttt{legal\_hard}) are excluded before train, validation, and held-out test construction to avoid distortion of the fitness signal~\citep{prbench}.

\subsection{LoCoMo}
\label{app:dataset-locomo}

LoCoMo~\citep{locomo} is a multi-session conversational question answering benchmark in which the agent must recover relevant facts from long dialogue histories.
The dataset contains extended dialogues with timestamped sessions, and the questions cover single-hop recall, temporal reasoning, causal reasoning, and cross-session multi-hop reasoning.
We use the provided conversation sessions as episodes and report token-level F1 together with an LLM-judge score.

\subsection{ALFWorld}
\label{app:dataset-alfworld}

ALFWorld~\citep{alfworld} evaluates embodied task completion in simulated household environments through text interaction.
The agent must complete goals such as finding, heating, cleaning, and placing objects.
We use expert trajectories from the training split as episodes and report success rate within a 50-step budget on both seen and unseen environment splits.

\subsection{HealthBench}
\label{app:dataset-healthbench}

HealthBench~\citep{healthbench} is a medical question answering benchmark with professional rubric-based evaluation.
Each sample includes a multi-turn clinical dialogue, an ideal completion, and structured pass or fail criteria written by medical professionals.
We evaluate two categories: health data interpretation and emergency referral recognition.
Because HealthBench does not provide a training split, we construct episode pools from a held-out portion of the benchmark data and keep evaluation queries disjoint from episode construction.
We filter by category first, then use 54\% of the category data for episode construction and reserve 100 examples for held-out testing.
We report rubric score, computed as the fraction of criteria marked positive by an LLM judge.

\subsection{PRBench}
\label{app:dataset-prbench}
\looseness=-1
PRBench~\citep{prbench} evaluates professional reasoning in legal analysis and financial valuation.
Each sample contains a task prompt, an expert reference, and an importance-weighted rubric.
Similar to HealthBench, PRBench does not provide a training split, so we construct episodes from a held-out portion and evaluate on disjoint query sets.
We filter out the hard subsets before splitting, use 40\% of the remaining data for episode construction, and reserve 50 examples for held-out testing.
For consistency, we apply the same rubric-based scoring protocol used for HealthBench to both PRBench splits in our evaluation.

\FloatBarrier

\section{Baseline Methods}
\label{app:baselines}

We compare \sysname{} against a no-memory control and nine memory-based baselines spanning retrieval-based systems, self-evolution systems, prompt-optimizing systems, and program-evolving systems.

\subsection{No Memory}
\label{app:baseline-no-memory}

The no-memory control answers each task directly without any external memory store or retrieved context during evaluation.

\subsection{Retrieval-Based Systems}
\label{app:baseline-retrieval}

These methods store raw observations and retrieve relevant context by similarity.
\textit{Vector Search} stores each observation in a vector collection and retrieves top-$k$ items by embedding cosine similarity, without task-specific parsing~\citep{rag}.
\textit{G-Memory}~\citep{alma} augments vector retrieval with a hierarchical graph over related tasks implemented with SQLite.
\textit{Mem0}~\citep{mem0} extracts atomic facts from observations and maintains them with LLM-driven add, update, and delete operations as new observations arrive.

\subsection{Self-Evolution Systems}
\label{app:baseline-self-evolution}

These methods distill reusable knowledge from experience instead of storing raw episodes directly.
\textit{Trajectory Retrieval}~\citep{alma} retrieves full episode trajectories by embedding similarity.
\textit{ReasoningBank}~\citep{ouyang2025reasoningbank} extracts key insights from each episode and retrieves them with similar task descriptions.
\textit{Dynamic Cheatsheet}~\citep{alma} maintains a single global summary that is iteratively rewritten after each new experience.

\subsection{Prompt-Optimizing Systems}
\label{app:baseline-prompt}

These methods improve agent behavior by evolving the system prompt.
\textit{GEPA}~\citep{gepa} optimizes prompts through reflective mutation with Pareto-based candidate selection.
We evaluate two variants: \textit{GEPA} and \textit{GEPA + Vector Search}.
The second variant adds a vector database to GEPA because the original GEPA setup does not include a persistent knowledge base, which can limit performance on memory-intensive tasks.

\subsection{Program-Evolving Systems}
\label{app:baseline-program}

\textit{MemEvolve}~\citep{memevolve} evolves a memory system as a program composed of encode, store, retrieve, and manage modules, which a coding agent rewrites across rounds.
We use its official implementation and the hyperparameters reported in the original paper, and we replace its optimization model with GPT-5.3-Codex to match \sysname{}.
On ALFWorld, MemEvolve stopped early after one (Unseen) or two (Seen) of its three rounds because no new system passed its own validator, and we report the best system found up to that point.

\section{Experiment Configuration}
\label{app:experiment-config}

\subsection{Computational Cost}
\label{app:cost}

\autoref{tab:cost} reports the computational cost of running \sysname{} evolution and baselines across all benchmarks.
We break down API calls by role: the \emph{task agent} (knowledge extraction, query generation, and answer generation), the \emph{reflector} (code mutation via LLM reflection), the \emph{judge} (per-criterion rubric scoring for HealthBench and PRBench), and the \emph{toolkit} (any LLM call issued from inside the memory pipeline at write or read time, including evolved \texttt{write()}/\texttt{read()} methods, baseline retrieval steps, and embedding-driven preprocessing helpers).
All experiments use \texttt{gpt-5.4-mini} as the task agent, judge, and toolkit model, and \texttt{gpt-5.3-codex} as the reflector model~\citep{openai2026gpt54mini,openai2026gpt53codex}.
Cost estimates use Azure OpenAI pricing: \$0.75/\$4.50 per 1M input/output tokens for \texttt{gpt-5.4-mini}, and \$1.75/\$14.00 per 1M input/output tokens for \texttt{gpt-5.3-codex}~\citep{openai2026gpt54mini,openai2026gpt53codex}.
Note that \texttt{gpt-5.3-codex} is a reasoning model; its output token counts include internal reasoning tokens, which are billed at the same rate as visible output tokens when computing the estimates below.
Token columns report GPT-family LLM tokens.
BGE-M3 embedding tokens used for vector search and subset selection are tracked separately from the GPT token and dollar estimates.

\begin{table}[!htbp]
\centering
\caption{\textbf{Computational cost of evolution runs and representative baselines.}
Rows report iteration count, API calls by role, GPT-family token counts, estimated dollar cost, and wall-clock time.
Wall-clock time includes environment interaction overhead for ALFWorld runs.}
\label{tab:cost}
\small
\setlength{\tabcolsep}{4pt}
\begin{tabular}{@{}ll r rrrr rr r r@{}}
\toprule
\textbf{Benchmark} & \textbf{Config} & \textbf{Iter.} & \textbf{Task} & \textbf{Reflect.} & \textbf{Judge} & \textbf{Toolkit} & \textbf{Input} & \textbf{Output} & \textbf{Cost} & \textbf{Time} \\
\midrule
LoCoMo        & \sysname{} & 20 & 5802 & 48 & --- & 135 & 4.9M & 4.2M & \$25.43 & 5.1h \\
              & Vector Search     &  0 &  200 &  0 & --- & 272 & 294K &  71K &  \$0.54 &  2m \\
              & No Memory       &  0 &  200 &  0 & --- & 272 & 219K &  64K &  \$0.45 &  1m \\
\midrule
ALFWorld      & \sysname{} & 20 & 3355 & 25 & --- & 443 & 7.5M & 1.4M & \$13.92 & 100h \\
(Unseen)      & Vector Search     &  0 &   80 &  0 & --- & 497 & 711K &  50K &  \$0.76 & 14m \\
              & No Memory       &  0 &   80 &  0 & --- & 497 & 708K &  50K &  \$0.76 & 11m \\
\midrule
ALFWorld      & \sysname{} & 20 & 2420 & 28 & --- & 494 & 5.6M & 1.1M & \$11.94 & 101h \\
(Seen)        & Vector Search     &  0 &   80 &  0 & --- & 497 & 692K &  49K &  \$0.74 & 11m \\
              & No Memory       &  0 &   50 &  0 & --- & 224 & 264K &  25K &  \$0.31 &  0m \\
\midrule
HB-Emergency  & \sysname{} & 20 & 4272 & 31 & 21370 & 2219 & 21M & 6.1M & \$47.20 & 4.3h \\
              & Vector Search     &  0 &   45 &  0 &  1102 &  423 & 830K & 145K &  \$1.28 &  6m \\
              & No Memory       &  0 &   54 &  0 &  1102 &  412 & 761K & 137K &  \$1.19 &  1m \\
\midrule
HB-DataTasks  & \sysname{} & 20 & 4204 & 24 & 15006 & 1575 & 18M & 6.3M & \$44.64 & 2.1h \\
              & Vector Search     &  0 &   24 &  0 &  1097 &  433 & 950K & 177K &  \$1.51 &  7m \\
              & No Memory       &  0 &   34 &  0 &  1105 &  423 & 836K & 171K &  \$1.40 &  2m \\
\midrule
PR-Legal      & \sysname{} & 20 & 3065 & 28 & 19984 & 1242 & 48M & 11M & \$90.19 & 15h \\
              & Vector Search     &  0 &   15 &  0 &   958 &  185 & 818K & 130K &  \$1.20 &  6m \\
              & No Memory       &  0 &   15 &  0 &   950 &  185 & 712K & 117K &  \$1.06 &  1m \\
\midrule
PR-Finance    & \sysname{} & 20 & 1795 & 31 & 18602 & 2497 & 47M & 10M & \$84.81 & 2.6h \\
              & Vector Search     &  0 &    8 &  0 &   845 &  212 & 740K & 153K &  \$1.24 &  6m \\
              & No Memory       &  0 &   13 &  0 &   895 &  207 & 646K & 139K &  \$1.11 & 11m \\
\bottomrule
\end{tabular}
\vspace{0.3em}

{\footnotesize Call counts include recorded API attempts.
Token totals sum provider usage fields when available; failed attempts without usage metadata contribute to call counts only.}
\end{table}

The total evolution cost ranges from \$12--90 per benchmark over 20 iterations, depending primarily on the evaluation protocol.
Rubric-graded benchmarks (HealthBench and PRBench) are substantially more expensive because each validation item requires multiple LLM judge calls for per-criterion scoring, making the judge the dominant cost component (60--80\% of total calls).
Token-F1 benchmarks (LoCoMo) and binary-success benchmarks (ALFWorld) avoid judge calls entirely, keeping evolution under \$26.
ALFWorld's long wall-clock time (100h) is dominated by TextWorld environment interaction; API latency contributes less.
Although the reflector makes only 24--48 calls per run, it uses the more expensive \texttt{gpt-5.3-codex} model (\$14/1M output tokens vs.\ \$4.50 for \texttt{gpt-5.4-mini}) and produces long code patches, so it accounts for 7--24\% of total cost depending on the benchmark.

Compared to baselines, the evolution overhead is a one-time cost: once the best program is found, inference uses the same number of calls as any baseline.
The per-query inference cost of an evolved program equals that of the corresponding baseline configuration (No Memory or Vector Search) plus any toolkit calls the evolved program makes, typically 0--2 additional LLM calls per query.

\FloatBarrier

\subsection{Hyperparameters}
\label{app:hyperparameters}

\autoref{tab:hyperparams} lists the key hyperparameters used across all experiments.

\begin{table}[!htbp]
\centering
\caption{\textbf{Hyperparameters for \sysname{} evolution.}
Rows list shared settings and benchmark-specific ranges used in the main experiments.}
\label{tab:hyperparams}
\small
\begin{tabular}{@{}ll@{}}
\toprule
\textbf{Parameter} & \textbf{Value} \\
\midrule
Evolution iterations         & 20 \\
Population seed programs     & 3 \\
Selection strategy           & Softmax ($\tau=0.15$) \\
Static validation subset     & 32--60 items ($k$-means clustering) \\
Rotating validation subset   & 5 items per iteration \\
Train-to-validation ratio    & 2 (train episodes = $2 \times$ validation items) \\
Embedding model (subset selection) & BGE-M3~\citep{bge_m3} \\
Reflector model              & GPT-5.3-Codex (reasoning: medium) \\
Task agent / toolkit model   & GPT-5.4-mini \\
Toolkit LLM call budget      & 1 call per \texttt{read}/\texttt{write} invocation \\
Compile-fix attempts ($k$)   & 3 \\
Evaluator thread pool        & 64 workers \\
LLM retry attempts           & 3 (exponential backoff) \\
\bottomrule
\end{tabular}
\end{table}

\subsection{Representative Subset Selection}
\label{app:representative-subset}

We construct representative subsets to reduce evaluation cost while preserving coverage of the full validation set.
For the static validation subset, we embed all validation samples, cluster them into $n$ groups using $k$-means~\citep{macqueen1967kmeans} with a fixed seed, and select the sample closest to each cluster centroid to form the fixed static validation set.
This procedure covers diverse regions of the validation distribution more effectively than random sampling under the same evaluation budget.
For the rotating validation subset used as reflection context, we re-cluster the remaining validation pool at each iteration with a different $k$-means seed (the iteration index added to a base seed), and select the sample closest to each new cluster centroid; varying the seed perturbs the cluster boundaries, so the chosen samples differ across iterations even though both clustering and centroid selection are deterministic given a seed.
The static subset determines fitness, while the rotating subset supplies diverse examples for mutation.

For episode subset selection, we use a facility-location objective to select $M$ training episodes that are relevant to the validation samples~\citep{nemhauser1978submodular}.
Given validation samples $V$ and candidate episodes $E$, we select a subset $S \subseteq E$ of size $M$ to maximize:
\begin{equation}
    \max_{S \subseteq E, |S|=M} \sum_{v \in V} \max_{e \in S} \operatorname{sim}(v, e),
\end{equation}
where $\operatorname{sim}(\cdot, \cdot)$ denotes embedding similarity.
We use greedy facility-location selection in the same embedding space~\citep{nemhauser1978submodular}.

\FloatBarrier

\section{Seed Programs}
\label{app:seeds}

The program pool is initialized with three structurally diverse seeds designed to cover different points in the memory-program search space.
All three share identical instruction constants but differ in storage strategy and retrieval logic.
\autoref{tab:seed-comparison} summarizes their key design choices.

\begin{table}[!htbp]
\centering
\caption{\textbf{Seed program comparison.}
Rows summarize each seed program's storage strategy, retrieval strategy, and read-time LLM use.}
\label{tab:seed-comparison}
\small
\begin{tabular}{@{}llll@{}}
\toprule
\textbf{Seed} & \textbf{Storage} & \textbf{Retrieval} & \textbf{LLM in \texttt{read}} \\
\midrule
Vector Search & ChromaDB chunks & Top-$k$ cosine similarity & No \\
LLM Summarizer & In-memory list & Concatenate + LLM summarize & Yes (1 call) \\
Experience Learner & Separate lesson/fact lists & Return all & No \\
\bottomrule
\end{tabular}
\end{table}

\subsection{Vector Search}
\label{app:seed-vector-search}

The Vector Search seed (\autoref{lst:seed-vector}) implements vanilla RAG~\citep{rag}: raw text is split into 500-character paragraph-aligned chunks and stored in a ChromaDB collection; \texttt{read()} retrieves the top-5 most similar documents by embedding cosine similarity.
No LLM call is used at retrieval time.

\subsection{LLM Summarizer}
\label{app:seed-llm-summarizer}

The LLM Summarizer seed (\autoref{lst:seed-llm}) stores raw text in an in-memory list without any preprocessing.
At retrieval time, all stored text is concatenated (up to 30{,}000 characters) and passed to the toolkit LLM alongside the query for query-focused summarization.
This seed tests whether neural synthesis at read time can compensate for the lack of structured indexing.

\subsection{Experience Learner}
\label{app:seed-experience-learner}

The Experience Learner seed (\autoref{lst:seed-experience}) extracts a general lesson and a specific fact from each observation, storing them in separate lists.
At retrieval time, it returns all stored lessons and facts (truncated to 500 characters each), ignoring the query entirely.
This seed explores whether dual-track extraction with full recall outperforms selective retrieval.

\begin{lstlisting}[style=evolvedcode, caption={Seed 1: Vector Search.}, label=lst:seed-vector]
class KnowledgeBase:
    """Vanilla RAG: store text chunks in ChromaDB, retrieve by semantic similarity."""
    def __init__(self, toolkit):
        self.toolkit = toolkit
        self.collection = toolkit.chroma.get_or_create_collection("knowledge")
        self._doc_id = 0

    def write(self, item: KnowledgeItem, raw_text: str) -> None:
        chunks = self._chunk(raw_text, max_chars=500)
        for chunk in chunks:
            self.collection.add(documents=[chunk], ids=[f"doc_{self._doc_id}"])
            self._doc_id += 1

    def read(self, query: Query) -> str:
        if self._doc_id == 0:
            return "No information stored."
        results = self.collection.query(
            query_texts=[query.query_text], n_results=min(5, self._doc_id))
        docs = results["documents"][0] if results["documents"] else []
        return "\n\n".join(docs)[:3000] if docs else "No relevant information found."
\end{lstlisting}

\begin{lstlisting}[style=evolvedcode, caption={Seed 2: LLM Summarizer.}, label=lst:seed-llm]
class KnowledgeBase:
    """LLM-powered query-focused summarization over stored raw texts."""
    def __init__(self, toolkit):
        self.toolkit = toolkit
        self.raw_texts: list[str] = []

    def write(self, item: KnowledgeItem, raw_text: str) -> None:
        self.raw_texts.append(raw_text)

    def read(self, query: Query) -> str:
        if not self.raw_texts:
            return "No information stored."
        combined = "\n\n".join(self.raw_texts)[:30000]
        messages = [{"role": "user", "content":
            f"Given the following query, summarize ONLY the relevant information "
            f"from the provided texts. Be concise and factual.\n\n"
            f"Query: {query.query_text}\n\nTexts:\n{combined}"}]
        try:
            result = self.toolkit.llm_completion(messages)
        except Exception:
            result = combined
        return result[:3000]
\end{lstlisting}

\begin{lstlisting}[style=evolvedcode, caption={Seed 3: Experience Learner.}, label=lst:seed-experience]
@dataclass
class KnowledgeItem:
    lesson_learned: str = field(
        metadata={"description": "A general lesson or pattern learned from the text"})
    fact_to_remember: str = field(
        metadata={"description": "A specific fact worth remembering from the text"})

class KnowledgeBase:
    """Experience-driven learner that stores lessons and facts, returns all on read."""
    def __init__(self, toolkit):
        self.toolkit = toolkit
        self.lessons: list[str] = []
        self.facts: list[str] = []

    def write(self, item: KnowledgeItem, raw_text: str) -> None:
        self.lessons.append(item.lesson_learned)
        self.facts.append(item.fact_to_remember)

    def read(self, query: Query) -> str:
        if not self.lessons and not self.facts:
            return "No information stored."
        lessons_text = "\n".join(self.lessons)[:500]
        facts_text = "\n".join(self.facts)[:500]
        return f"Lessons:\n{lessons_text}\n\nFacts:\n{facts_text}"[:3000]
\end{lstlisting}

\FloatBarrier

\section{Evolved Programs}
\label{app:evolved-programs}

We present the best evolved programs for two representative benchmarks, selected to illustrate the structural diversity described in Section~\ref{sec:results}.
Full source code for all seven benchmark configurations is included in the released experiment artifacts under each run's \texttt{programs/} directory.

\subsection{ALFWorld: Deterministic Action Cache}
\label{app:evolved-alfworld}

The best ALFWorld program (\autoref{lst:alf-write}, \autoref{lst:alf-read}) constructs a deterministic action cache using SQLite.
It stores structured fields extracted from expert demonstrations---target object, destination, required state change (clean/cool/heat/examine/toggle), action hints, and failure modes---and retrieves them via a weighted scoring function that combines token overlap with exact-match bonuses for object, location, and state.
The program includes a \texttt{\_canonical\_state()} normalizer that maps diverse surface forms (``rinse'', ``wash'' $\to$ ``clean''; ``chill'', ``cold'' $\to$ ``cool'') to canonical labels, and a fallback parser that extracts task metadata from raw text when the LLM's structured extraction is sparse.
A single LLM call at read time synthesizes the top-6 retrieved memories into concise step-by-step guidance for the task agent to follow.
This description corresponds to the best ALFWorld Unseen program.
The best ALFWorld Seen program uses the same SQLite-backed design family, but its \texttt{read()} method returns deterministic ranked guidance without a toolkit LLM call.

The key architectural properties are:
\begin{itemize}
\item \textbf{SQLite-only storage}, with zero vector retrieval (ChromaDB unused).
\item \textbf{Canonical state normalization} unifies synonymous state descriptions (6 canonical states).
\item \textbf{Keyword-based scoring} with exact-match bonuses (+6 for object, +5 for location, +5 for state) and recency tiebreaker for stable ranking.
\item \textbf{Schema}: 8 structured fields per memory item (task\_summary, task\_type, target\_object, target\_location, required\_state, action\_hint, failure\_mode, keywords).
\end{itemize}

\begin{lstlisting}[style=evolvedcode, caption={ALFWorld evolved \texttt{write()} --- canonical state extraction and fallback parsing.}, label=lst:alf-write]
def _canonical_state(self, text):
    t = (text or "").lower()
    if any(k in t for k in ["rinse", "wash", "clean"]): return "clean"
    if any(k in t for k in ["cool", "chill", "cold", "refrigerat"]): return "cool"
    if any(k in t for k in ["heat", "warm", "hot"]): return "heat"
    if any(k in t for k in ["examine", "inspect", "look at"]): return "examine"
    if any(k in t for k in ["toggle", "turn on", "light"]): return "toggle"
    return ""

def write(self, item, raw_text):
    task_summary = self._clean(item.task_summary)
    task_type = self._clean(item.task_type)
    target_object = self._clean(item.target_object)
    target_location = self._clean(item.target_location)
    action_hint = self._clean(item.action_hint)
    failure_mode = self._clean(item.failure_mode)
    required_state = self._canonical_state(item.required_state) or \
        self._canonical_state(f"{task_summary} {task_type} {raw_text}")
    # Fallback: extract from raw task metadata when LLM fields are sparse
    if not target_object:
        m = re.search(r"object_target:\s*([A-Za-z0-9_]+)", raw_text)
        if m: target_object = m.group(1)
    kw = set()
    for part in [task_summary, task_type, target_object, target_location]:
        kw |= self._tokenize(part)
    for k in item.keywords: kw |= self._tokenize(k)
    self.db.execute("INSERT INTO memories (...) VALUES (...)",
        (task_summary, task_type, target_object, target_location,
         required_state, action_hint, failure_mode, json.dumps(sorted(kw)),
         raw_text[:2000]))
\end{lstlisting}

\begin{lstlisting}[style=evolvedcode, caption={ALFWorld evolved \texttt{read()} --- weighted scoring with exact-match bonuses.}, label=lst:alf-read]
def read(self, query):
    rows = self.db.execute("SELECT id, task_summary, task_type, "
        "target_object, target_location, required_state, action_hint, "
        "failure_mode, keywords_json FROM memories").fetchall()
    q_obj = self._clean(query.target_object)
    q_loc = self._clean(query.target_location)
    q_state = self._canonical_state(query.required_state)
    q_tokens = self._tokenize(query.request) | self._tokenize(q_obj) | self._tokenize(q_loc)
    scored = []
    for row in rows:
        (row_id, _, _, target_object, target_location,
         required_state, _, _, keywords_json) = row
        mem_kw = set(json.loads(keywords_json or "[]"))
        score = len(q_tokens & mem_kw)                        (*@\ding{172}@*)
        if q_obj == (target_object or ""): score += 6         (*@\ding{173}@*)
        if q_loc == (target_location or ""): score += 5       (*@\ding{174}@*)
        if q_state == (required_state or ""): score += 5      (*@\ding{175}@*)
        score += min(row_id, 1000) / 100000.0                 (*@\ding{176}@*)
        scored.append((score, row))
    selected = sorted(scored, reverse=True, key=lambda x: x[0])[:6]
    # One LLM call: synthesize top memories into step-by-step guidance
    synthesized = self.toolkit.llm_completion([...])           (*@\ding{177}@*)
    return synthesized[:3000]
\end{lstlisting}
\vspace{-0.5em}
{\scriptsize \ding{172}~Token overlap \quad \ding{173}~Exact object match \quad \ding{174}~Exact location match \quad \ding{175}~State match \quad \ding{176}~Recency tiebreaker \quad \ding{177}~LLM synthesis}

\subsection{LoCoMo: Multi-Signal Episodic Index}
\label{app:evolved-locomo}

The best LoCoMo program builds a hybrid memory combining SQLite and ChromaDB (\autoref{lst:loco-schema}, \autoref{lst:loco-read}).
Each observation is parsed into 7 structured metadata fields (participants, organizations, activities, key facts, relation facts, named entities) and stored in both a relational table and a vector collection.
At retrieval time, the program fuses semantic similarity scores from ChromaDB with lexical overlap scores from SQLite, applies person-focused boosting when the query targets a specific individual, limits per-source diversity (at most 2 chunks from any single dialogue), and separately ranks extracted facts by query relevance.
The final output combines top-ranked candidate facts and relevant text excerpts for answer generation.

The key architectural properties are:
\begin{itemize}
\item \textbf{425 source lines}, with the largest hybrid retrieval procedure among the best programs.
\item \textbf{Dual storage}: SQLite for structured metadata + ChromaDB for semantic search.
\item \textbf{7 metadata fields} per item including relation facts (subject-verb-object triples).
\item \textbf{Source diversity cap}: maximum 2 chunks per source document, preventing any single conversation from dominating retrieval for a query.
\item \textbf{Two-tier output}: candidate facts (direct answer candidates) and relevant excerpts (contextual evidence for grounded answer generation in responses).
\end{itemize}

\begin{lstlisting}[style=evolvedcode, caption={LoCoMo evolved schema --- 7 structured metadata fields.}, label=lst:loco-schema]
@dataclass
class KnowledgeItem:
    summary: str = field(metadata={"description": "Short summary"})
    participants: list[str] = field(
        metadata={"description": "People explicitly discussed in the text"})
    organizations: list[str] = field(
        metadata={"description": "Organizations, groups, or institutions"})
    activities: list[str] = field(
        metadata={"description": "Activities, hobbies, events mentioned"})
    key_facts: list[str] = field(
        metadata={"description": "Atomic factual statements useful for QA"})
    relation_facts: list[str] = field(
        metadata={"description": "Subject-verb-object facts"})
    named_entities: list[str] = field(
        metadata={"description": "Exact named items: orgs, games, places"})
\end{lstlisting}

\begin{lstlisting}[style=evolvedcode, caption={LoCoMo evolved \texttt{read()} excerpt --- hybrid scoring and source diversity.}, label=lst:loco-read]
def read(self, query):
    # Phase 1: Semantic retrieval from ChromaDB
    results = self.collection.query(query_texts=[query_text], n_results=24)
    ids = results["ids"][0]; dists = results["distances"][0]
    for rank, (doc_id, dist) in enumerate(zip(ids, dists)):
        semantic_scores[doc_id] = 1.25 / (rank + 1) + 1.0 / (1 + dist)

    # Phase 2: Lexical scoring from SQLite
    for row in rows:
        lexical = self._overlap_score(query_tokens, doc_tokens)
        person_boost = 1.2 if focus_person in participants else 0.0
        scores[doc_id] += 2.1 * lexical + 1.0 * info_overlap + person_boost

    # Phase 3: Source diversity enforcement
    source_counts = collections.Counter()
    for doc_id, _ in ranked_pairs:
        source_id = row_by_id[doc_id]["source_id"]
        if source_counts[source_id] >= 2: continue     (*@\ding{172}@*)
        ranked_ids.append(doc_id)
        source_counts[source_id] += 1

    # Phase 4: Separate fact ranking
    for doc_id in ranked_ids:
        for fact in relation_facts + key_facts:
            fscore = overlap(query_tokens, fact_tokens)
            if focus_person in fact: fscore += 0.5      (*@\ding{173}@*)
            fact_candidates.append((fscore, fact))

    # Output: candidate facts + relevant excerpts
    return "Candidate facts:\n..." + "\nExcerpts:\n..."  (*@\ding{174}@*)
\end{lstlisting}
\vspace{-0.5em}
{\scriptsize \ding{172}~Max 2 chunks per source \quad \ding{173}~Person-focused boost \quad \ding{174}~Two-tier output format}

\subsection{HealthBench and PRBench Program Sketches}
\label{app:evolved-health-pr}

The HealthBench and PRBench programs evolve toward criterion-aware retrieval rather than broad episodic recall.
The HealthBench variants use SQLite tables keyed by clinical topic, urgency signals, patient attributes, and rubric-relevant safety notes; the Data Tasks variant also uses ChromaDB for semantic retrieval.
At read time, they prioritize memories that match the current clinical intent and then use one toolkit LLM call to synthesize concise guidance for the task agent.
This structure reflects the benchmark's grading protocol: answers must cover medically relevant criteria while avoiding unsafe or unsupported advice.

The PRBench variants emphasize domain-specific decomposition.
The legal program stores issue statements, governing rules, factual predicates, counterarguments, and citation-like anchors, then retrieves by matching the requested legal theory and factual pattern.
The finance program stores assumptions, valuation drivers, comparable-company cues, risk factors, and calculation notes, then combines structured scoring with vector retrieval for rubric-targeted evidence.
These programs illustrate the same pattern as ALFWorld and LoCoMo: each task induces a different schema and retrieval procedure.

\subsection{Cross-Benchmark Structural Comparison}
\label{app:evolved-comparison}

\autoref{tab:evolved-comparison} summarizes the architectural differences across all four benchmark domains, illustrating how reflective code evolution discovers structurally distinct solutions for each task.

\begin{table}[!htbp]
\centering
\caption{\textbf{Structural comparison of best evolved programs across benchmark configurations.}
Rows report program size, storage backends, read-time LLM use, schema size, and test score.}
\label{tab:evolved-comparison}
\small
\setlength{\tabcolsep}{3pt}
\begin{tabular}{@{}lcccccc@{}}
\toprule
\textbf{Benchmark} & \textbf{Lines} & \textbf{SQLite} & \textbf{ChromaDB} & \textbf{LLM in \texttt{read}} & \textbf{Schema fields} & \textbf{Test} \\
\midrule
LoCoMo              & 425  & Yes & Yes & No  & 7  & 0.459 \\
ALFWorld (Unseen)   & 276  & Yes & No  & Yes & 8  & 0.881 \\
ALFWorld (Seen)     & 441  & Yes & No  & No  & 8  & 0.700 \\
HB-Emergency        & 393  & Yes & No  & Yes & 7  & 0.493 \\
HB-Data             & 769  & Yes & Yes & Yes & 12 & 0.390 \\
PR-Legal            & 1184 & Yes & Yes & Yes & 22 & 0.660 \\
PR-Finance          & 509  & Yes & Yes & Yes & 10 & 0.586 \\
\bottomrule
\end{tabular}
\vspace{0.3em}

{\footnotesize Lines count the released best-program source files.}
\end{table}

\FloatBarrier

\section{Prompt Templates and Mutation Interface}
\label{app:prompts}

\looseness=-1
This section documents the key LLM prompts used in \sysname{}.
We categorize prompts by their role in the system: task agent prompts (used during evaluation) and reflector prompts (used during evolution).

\subsection{Task Agent Prompts}
\label{app:task-agent-prompts}

The task agent interacts with evolved memory programs through three fixed prompt templates.
These templates are parameterized by the program's \texttt{INSTRUCTION\_*} constants, meaning evolution can steer the agent's behavior by modifying these constants without changing the prompt structure.

\paragraph{Knowledge item generation.}
Given raw text and a \texttt{KnowledgeItem} schema, the task agent extracts structured information for a \texttt{write} call:
\begin{lstlisting}[style=evolvedcode, basicstyle=\ttfamily\scriptsize]
{INSTRUCTION_KNOWLEDGE_ITEM}

Text: {raw_text}

The KnowledgeItem must conform to this schema:
{schema}

Output ONLY a valid JSON object matching the schema fields. No explanation.
\end{lstlisting}

\paragraph{Query generation.}
Given a question and a \texttt{Query} schema, the task agent formulates a retrieval query for a \texttt{read} call:
\begin{lstlisting}[style=evolvedcode, basicstyle=\ttfamily\scriptsize]
{INSTRUCTION_QUERY}

Question: {question}

The query must be a JSON object matching this schema:
{schema}

Respond with the JSON only.
\end{lstlisting}

\paragraph{Answer generation.}
Given retrieved memory, the task agent generates an answer.
The \texttt{ALWAYS\_ON\_KNOWLEDGE} constant, if non-empty, is prepended to the retrieved content:
\begin{lstlisting}[style=evolvedcode, basicstyle=\ttfamily\scriptsize]
<retrieved_memory>
{ALWAYS_ON_KNOWLEDGE}

{retrieved}
</retrieved_memory>

{INSTRUCTION_RESPONSE}
\end{lstlisting}

\subsection{Reflector Prompt}
\label{app:reflector-prompt}

The reflector receives the current program, its evaluation score, underperforming cases, and optional context (reference programs, lineage history), and outputs a V4A patch to improve the program.
The complete prompt template is shown below.
Sections in \texttt{\{braces\}} are populated dynamically at each iteration; optional sections (lineage log, write examples, success cases, reference programs) are included only when available for that iteration.

\paragraph{Interface specification.}
The following specification is provided to the reflector to define the \texttt{KnowledgeBase} API:

\begin{lstlisting}[style=evolvedcode, basicstyle=\ttfamily\scriptsize]
You are designing a Knowledge Base Program that implements three classes:

1. **KnowledgeItem** (dataclass): Defines what information is captured
   as knowledge items when writing to the knowledge base.
   - Must be a @dataclass with typed fields
   - An external LLM will populate instances by generating JSON matching
     your field definitions
   - **Field types MUST be JSON-compatible**: use only str, int, float,
     bool, list[str], Optional[str]
   - Do NOT use datetime, tuple, bytes, or custom objects
   - Use `field(metadata={"description": "..."})` to describe fields

2. **Query** (dataclass): Defines what parameters are used when reading
   from the knowledge base.
   - Same constraints as KnowledgeItem

3. **KnowledgeBase** (class): The core knowledge base system.
   - `__init__(self, toolkit)`: Receives a Toolkit with:
     - `toolkit.db`: sqlite3.Connection (in-memory SQLite)
     - `toolkit.chroma`: chromadb ephemeral client
     - `toolkit.llm_completion(messages, **kwargs) -> str`: LLM for
       reasoning, summarization, and information extraction
       (1 call per write/read invocation)
     - `toolkit.logger.debug(message)`: Debug logging
   - `write(self, item: KnowledgeItem, raw_text: str) -> None`
   - `read(self, query: Query) -> str`

Allowed imports: json, re, math, hashlib, collections, dataclasses,
typing, datetime, textwrap, sqlite3, chromadb

## Runtime Constraints
- `read()` output limit: at most 3000 characters.
- `write()` / `read()` timeout: 60 seconds each.
- `toolkit.llm_completion()` budget: at most 1 LLM call per
  `write()` or `read()` invocation. The budget resets before each call.

## Instruction Constants (required)
Four module-level string constants:
- INSTRUCTION_KNOWLEDGE_ITEM: What to extract and how to structure it.
- INSTRUCTION_QUERY: How to formulate retrieval queries.
- INSTRUCTION_RESPONSE: Answer format, length, and style.
- ALWAYS_ON_KNOWLEDGE: Persistent context injected into every task
  agent prompt. Can be empty.
\end{lstlisting}

\paragraph{Patch format specification.}
The reflector is instructed to output changes in V4A patch format:

\begin{lstlisting}[style=evolvedcode, basicstyle=\ttfamily\scriptsize]
Before the patch, output a commit message summarizing your changes:

*** Commit Message
Title: <one-line summary of what you changed and why>
- <root cause / diagnosis>
- <what you changed>

Then output your changes as a V4A patch.

IMPORTANT: You MUST output the exact markers `*** Begin Patch` and
`*** End Patch` on their own lines. Do NOT wrap them in code fences.

Format:
*** Begin Patch
*** Update File: program.py
@@ <optional context hint>
 context line (1-2 lines before change)
-removed line
+added line
 context line (1-2 lines after change)
*** End Patch

Rules:
- Lines prefixed with `-` are removed, `+` are added,
  ` ` (space) are unchanged context.
- Include 1-2 context lines before and after each change.
- Multiple hunks are allowed within one `*** Update File` block.
\end{lstlisting}

\paragraph{Main reflection prompt.}
The following is the complete prompt template sent to the reflector LLM for each mutation step in evolution.

\begin{lstlisting}[style=evolvedcode, basicstyle=\ttfamily\scriptsize]
You are an expert Python programmer specializing in knowledge base
system design.

Your task: Given a Knowledge Base Program, its evaluation score, and
underperforming cases, identify the root cause of each low score and
improve the program. Improvements are two-fold:
(A) **Prompt Optimization** -- tune the four instruction constants
    (especially ALWAYS_ON_KNOWLEDGE) to steer the task agent's
    behavior, and
(B) **Memory Design** -- improve the KnowledgeItem/Query schemas and
    KnowledgeBase storage/retrieval logic.
Both dimensions matter and should be considered together.

<interface_spec>
{KB_INTERFACE_SPEC}
</interface_spec>

<rules>
1. Output your diagnosis first, then your changes as a patch.
2. The code must define exactly three classes (KnowledgeItem, Query,
   KnowledgeBase) and four module-level string constants
   (INSTRUCTION_KNOWLEDGE_ITEM, INSTRUCTION_QUERY,
   INSTRUCTION_RESPONSE, ALWAYS_ON_KNOWLEDGE).
3. KnowledgeBase.__init__ must accept `toolkit`; write takes a
   KnowledgeItem and a raw_text string; read takes a Query and
   returns str.
4. `read()` must return at most 3000 characters.
5. Keep it simple. Make minimal changes that generalize beyond the
   specific cases shown -- no hardcoded word lists or
   case-specific pattern rules.
6. **Prompt Optimization**: Update INSTRUCTION_* to steer the task
   LLM's output format. Update ALWAYS_ON_KNOWLEDGE with domain
   strategies, heuristics, and behavioral rules the task agent
   should always follow -- this constant is injected into EVERY
   task agent action/decision prompt and is often the
   highest-leverage change. Study the <model_generation> transcripts
   in the underperforming cases to identify agent behavioral
   patterns (e.g., looping, inefficient exploration, wrong object
   selection) that ALWAYS_ON_KNOWLEDGE can fix.
7. **Memory Design**: Improve KnowledgeItem/Query field schemas and
   KnowledgeBase read()/write() logic to store and retrieve more
   useful information for the task agent.
8. Add clear comments explaining WHY each part of the code works
   the way it does -- this helps future iterations understand and
   preserve your design decisions.
</rules>

<patch_format>
{PATCH_FORMAT_SPEC}
</patch_format>

<current_program iteration="{iteration}">
```python
{code}
```
</current_program>

<evaluation_score>{score}</evaluation_score>

{lineage_section}
{train_section}
{memory_debug_logs}
{success_section}
{reference_section}

The following cases show poor performance on the validation set after
memory has been written. Each case contains the full
retrieval-and-answer conversation trajectory.

<underperforming_cases>
<case id="1">
<question>{question}</question>
<rationale>{expected_answer}</rationale>
<model_generation>{model_output}</model_generation>
<score>{case_score}</score>
<conversation>
  [user]: {query_generation_prompt}
  [assistant]: {query_json}
  [user]: <retrieved_memory>...</retrieved_memory> {instruction}
  [assistant]: {answer}
</conversation>
</case>
...
</underperforming_cases>

<task>
1. Diagnose why these cases scored low -- examine both the retrieval
   conversation AND the <model_generation> transcript for agent
   behavioral issues.
2. Propose improvements along two dimensions:
   (A) **Prompt Optimization**: How should INSTRUCTION_* and
       ALWAYS_ON_KNOWLEDGE change to steer the task agent better?
   (B) **Memory Design**: How should the schemas or
       storage/retrieval logic change to provide more useful
       information?
3. Output your changes as a patch.
</task>
\end{lstlisting}

\paragraph{Optional sections.}
The following sections are conditionally included when their data is available:

\begin{itemize}
\item \textbf{Lineage log}: Evolution history of the current program's lineage (ancestors, children, regression markers), formatted as commit-style entries with delta scores.
Regression markers (\texttt{$\leftarrow$ REGRESSION}) flag changes that hurt performance, instructing the reflector not to repeat them.

\item \textbf{Write examples}: Sample knowledge ingestion trajectories showing how the external LLM generates knowledge items from raw document text and how \texttt{write()} is called.

\item \textbf{Success cases}: Cases where the current program performed well, instructing the reflector to preserve the behavior that makes these work.

\item \textbf{Reference programs}: Higher- or lower-scoring programs from the population, with instructions to study which design patterns (e.g., use of \texttt{llm\_completion}, ChromaDB vs.\ SQLite, schema granularity) correlate with scores within the pool.

\item \textbf{Memory debug logs}: Outputs of \texttt{toolkit.logger.debug()} calls within \texttt{write()} and \texttt{read()}, providing visibility into program execution.
\end{itemize}

\paragraph{Underperforming case selection.}
At each iteration, 2 failed cases are sampled from the rotating validation set using the Efraimidis--Spirakis weighted sampling algorithm with weight $w_i = 1 - \text{score}_i$, biasing selection toward lower-scoring cases while maintaining diversity~\citep{efraimidis2006weighted}.

\subsection{Compile-Fix Prompt}
\label{app:compile-fix-prompt}

When a mutated program fails to compile or violates runtime constraints, a compile-fix prompt is sent to the reflector with the failing program and error details:
\begin{lstlisting}[style=evolvedcode, basicstyle=\ttfamily\scriptsize]
You are an expert Python programmer. A Knowledge Base Program failed
to compile or run. Fix the error and output your fix as a patch.

{KB_INTERFACE_SPEC}

## Failing Code
```python
{code}
```

## Error
**{error_type}**: {error_details}

Fix the error and output your fix as a patch.
\end{lstlisting}
Up to $k$ compile-fix iterations are attempted before the mutation is discarded.

\FloatBarrier

\section{Evaluation Protocols}
\label{app:eval-protocols}

Each benchmark uses a task-specific evaluation metric.
All metrics return a score in $[0, 1]$; the fitness function $J(P)$ averages these scores across the evaluation set.

\subsection{Token F1 (LoCoMo)}
\label{app:metric-token-f1}

Following \citet{squad}, we compute token-level F1 between the model output and the expected answer.
Both strings are lowercased, articles (``a'', ``an'', ``the'') are removed, and all punctuation is stripped.
Precision and recall are computed over the resulting token multisets using the standard formula below:
\begin{equation}
F_1 = \frac{2 \cdot |\text{out} \cap \text{exp}|}{|\text{out}| + |\text{exp}|}.
\end{equation}
If both token sets are empty, the score is 1.0; if exactly one is empty, the score is 0.0.

\paragraph{LLM-Judge Score (LoCoMo).}
The L-J column in Table~\ref{tab:main-results} reports a complementary LLM-as-judge metric that allows for paraphrastic answers that token F1 would penalize.
A separate judge call grades each prediction against the gold answer and returns a binary correctness label, which we then average over the test set.
The judge uses the same task agent model (GPT-5.4-mini) and the prompt template below; the score is $1.0$ if the response parses as the integer $1$ and $0.0$ otherwise.
\begin{lstlisting}[style=evolvedcode, basicstyle=\ttfamily\scriptsize]
You are a strict judge. Determine if the output answers the question
correctly based on the expected answer. Reply ONLY with 1 (correct)
or 0 (incorrect).

Expected answer: <<expected>>
Actual output:   <<output>>

Score (0 or 1):
\end{lstlisting}

\subsection{Binary Success (ALFWorld)}
\label{app:metric-binary-success}

The agent interacts with the TextWorld environment for up to 50 steps~\citep{textworld,alfworld}.
A score of 1.0 is assigned if the task is completed within the step budget; 0.0 otherwise.

\subsection{Rubric-Based Scoring (HealthBench, PRBench)}
\label{app:metric-rubric}

Each evaluation item includes a structured rubric with multiple criteria, each assigned a point value (positive for desirable traits, negative for undesirable ones).
An LLM judge independently grades each criterion by examining the conversation and the last assistant response, returning a JSON object with an \texttt{explanation} and a boolean \texttt{criteria\_met} field.
For negative criteria (e.g., ``Is overly verbose''), a good response should be graded as \texttt{false} (the undesirable trait is not present).
The score is computed as:
\begin{equation}
\text{score} = \text{clip}\!\left(\frac{\sum_{i:\, \text{met}_i} p_i}{\sum_{i:\, p_i > 0} p_i},\; 0,\; 1\right),
\end{equation}
where $p_i$ is the point value for criterion $i$ and $\text{met}_i$ indicates whether the criterion was judged as met.
Negative criteria that are met subtract from the numerator.
This protocol follows the official HealthBench and PRBench evaluation methodology~\citep{healthbench,prbench}.
The LLM judge uses the following prompt template for each criterion during grading:

\begin{lstlisting}[style=evolvedcode, basicstyle=\ttfamily\scriptsize]
Your job is to look at a conversation and a rubric item, and score
the last turn (the last assistant response) on how well it follows
the rubric item.

# Conversation
<<conversation>>

# Rubric item
<<rubric_item>>

# Instructions
Return a json object with fields: "explanation" and "criteria_met".
- "explanation": why the response does or does not meet the criteria.
- "criteria_met": boolean. If a rubric item has multiple criteria,
  all must be met to return true.
- Exception: if criteria says "such as" or "for example", the
  response need not include all listed examples.
- For negative criteria (undesirable traits), return whether the
  criteria IS met (true = bad behavior present), not whether the
  response is good.
\end{lstlisting}

\section{Extended Results}
\label{app:extended-results}

\subsection{Complete Per-Category Results}
\label{app:per-category}

Table~\ref{tab:per-category} in the main paper reports per-category results for ALFWorld Unseen and LoCoMo with selected baselines.
Here we provide the complete per-category breakdown for all benchmarks and all baselines in the appendix tables below.

\begin{table}[!htbp]
\centering
\caption{\textbf{Complete per-category results for ALFWorld Unseen.}
Columns report success rate by task type.
Best per column appears in \textbf{bold} for each reported task category.}
\label{tab:per-cat-alf-unseen}
\small
\setlength{\tabcolsep}{4pt}
\begin{tabular}{@{}l cccccc c@{}}
\toprule
& \textbf{Simple} & \textbf{Movable} & \textbf{Clean} & \textbf{Light} & \textbf{Cool} & \textbf{Heat} & \textbf{2-Obj} \\
& \footnotesize$(n{=}8)$ & \footnotesize$(n{=}1)$ & \footnotesize$(n{=}11)$ & \footnotesize$(n{=}1)$ & \footnotesize$(n{=}7)$ & \footnotesize$(n{=}6)$ & \footnotesize$(n{=}8)$ \\
\midrule
No Memory            & 0.88 & \textbf{1.00} & 0.82 & 0.00 & 0.57 & 0.83 & 0.62 \\
Vector Search        & 0.62 & \textbf{1.00} & 0.64 & 0.00 & 0.71 & 0.67 & 0.62 \\
G-Memory             & \textbf{1.00} & \textbf{1.00} & 0.55 & \textbf{1.00} & 0.29 & 0.83 & 0.75 \\
Mem0                 & 0.88 & \textbf{1.00} & 0.64 & 0.00 & 0.57 & \textbf{1.00} & 0.75 \\
Trajectory Retrieval & 0.75 & \textbf{1.00} & 0.73 & 0.00 & 0.71 & 0.67 & 0.75 \\
ReasoningBank        & \textbf{1.00} & \textbf{1.00} & 0.55 & \textbf{1.00} & 0.57 & 0.83 & 0.75 \\
Dynamic Cheatsheet   & \textbf{1.00} & \textbf{1.00} & 0.36 & \textbf{1.00} & 0.00 & \textbf{1.00} & 0.75 \\
GEPA                 & 0.88 & \textbf{1.00} & \textbf{1.00} & 0.00 & \textbf{0.86} & \textbf{1.00} & 0.62 \\
GEPA + Vector Search & 0.88 & \textbf{1.00} & \textbf{1.00} & 0.00 & 0.57 & \textbf{1.00} & 0.62 \\
\rowcolor{green!10}
\sysname{}           & 0.88 & \textbf{1.00} & 0.91 & \textbf{1.00} & \textbf{1.00} & 0.83 & 0.75 \\
\bottomrule
\end{tabular}
\end{table}

\begin{table}[!htbp]
\centering
\caption{\textbf{Complete per-category results for ALFWorld Seen.}
Columns report success rate by task type.
Best per column appears in \textbf{bold} for each reported task category.}
\label{tab:per-cat-alf-seen}
\small
\setlength{\tabcolsep}{4pt}
\begin{tabular}{@{}l cccccc c@{}}
\toprule
& \textbf{Simple} & \textbf{Clean} & \textbf{Light} & \textbf{Cool} & \textbf{Heat} & \textbf{2-Obj} & \textbf{Overall} \\
& \footnotesize$(n{=}2)$ & \footnotesize$(n{=}12)$ & \footnotesize$(n{=}2)$ & \footnotesize$(n{=}8)$ & \footnotesize$(n{=}7)$ & \footnotesize$(n{=}15)$ & \footnotesize$(n{=}50)$ \\
\midrule
No Memory            & \textbf{1.00} & 0.33 & \textbf{1.00} & 0.50 & 0.43 & 0.87 & 0.640 \\
Vector Search        & \textbf{1.00} & 0.42 & \textbf{1.00} & \textbf{1.00} & 0.29 & 0.87 & 0.720 \\
G-Memory             & \textbf{1.00} & 0.33 & \textbf{1.00} & 0.12 & 0.29 & 0.87 & 0.560 \\
Mem0                 & \textbf{1.00} & 0.25 & \textbf{1.00} & 0.62 & 0.43 & 0.87 & 0.640 \\
Trajectory Retrieval & \textbf{1.00} & 0.58 & \textbf{1.00} & 0.88 & 0.43 & \textbf{0.93} & 0.780 \\
ReasoningBank        & \textbf{1.00} & 0.33 & \textbf{1.00} & 0.25 & 0.43 & 0.80 & 0.580 \\
Dynamic Cheatsheet   & \textbf{1.00} & 0.17 & \textbf{1.00} & 0.38 & 0.00 & 0.73 & 0.480 \\
GEPA                 & \textbf{1.00} & 0.50 & \textbf{1.00} & 0.88 & 0.43 & 0.80 & 0.720 \\
GEPA + Vector Search & \textbf{1.00} & \textbf{0.92} & \textbf{1.00} & \textbf{1.00} & 0.29 & 0.80 & \textbf{0.820} \\
\rowcolor{green!10}
\sysname{}           & \textbf{1.00} & 0.42 & \textbf{1.00} & 0.75 & \textbf{0.57} & 0.80 & 0.700 \\
\bottomrule
\end{tabular}
\vspace{0.3em}

{\footnotesize \textbf{Bold} marks the best per category; GEPA+VS has the highest overall score (0.820).}
\end{table}

\begin{table}[!htbp]
\centering
\caption{\textbf{Complete per-category results for LoCoMo.}
Columns report token F1 by question category: single-hop recall, temporal reasoning, causal reasoning, and open-domain.
Best per column appears in \textbf{bold} for each reported question category.}
\label{tab:per-cat-locomo}
\small
\setlength{\tabcolsep}{5pt}
\begin{tabular}{@{}l cccc c@{}}
\toprule
& \textbf{Single-hop} & \textbf{Temporal} & \textbf{Causal} & \textbf{Open-domain} & \textbf{Overall} \\
& \footnotesize$(n{=}18)$ & \footnotesize$(n{=}18)$ & \footnotesize$(n{=}9)$ & \footnotesize$(n{=}55)$ & \\
\midrule
No Memory            & 0.02 & 0.00 & 0.11 & 0.04 & 0.036 \\
Vector Search        & 0.19 & 0.29 & 0.38 & 0.25 & 0.256 \\
G-Memory             & 0.23 & 0.12 & 0.31 & 0.24 & 0.224 \\
Mem0                 & \textbf{0.33} & 0.26 & 0.37 & 0.43 & 0.373 \\
Trajectory Retrieval & 0.18 & \textbf{0.33} & 0.34 & 0.28 & 0.276 \\
ReasoningBank        & 0.25 & 0.07 & 0.34 & 0.19 & 0.194 \\
Dynamic Cheatsheet   & 0.14 & 0.00 & 0.22 & 0.14 & 0.124 \\
GEPA                 & 0.09 & 0.06 & 0.24 & 0.15 & 0.132 \\
GEPA + Vector Search & 0.16 & \textbf{0.38} & \textbf{0.39} & 0.31 & 0.300 \\
\rowcolor{green!10}
\sysname{}           & \textbf{0.33} & 0.19 & 0.35 & \textbf{0.51} & \textbf{0.459} \\
\bottomrule
\end{tabular}
\end{table}

HealthBench and PRBench each evaluate on single-category splits (emergency referrals, health data interpretation, legal reasoning, and financial reasoning).
Their per-category scores equal the overall scores reported in Table~\ref{tab:main-results}; we omit separate per-category tables for these benchmarks.

\subsection{Multi-Seed Stability Results}
\label{app:stability}

Table~\ref{tab:stability} in the main paper reports summary statistics (mean, standard deviation, coefficient of variation) across five evolution seeds.
Here we provide the full per-seed results.

\begin{table}[!htbp]
\centering
\caption{\textbf{Per-seed test scores across stability experiments.}
Best Baseline denotes the strongest fixed baseline.}
\label{tab:stability-full}
\small
\setlength{\tabcolsep}{4pt}
\begin{tabular}{@{}l ccccc ccc c@{}}
\toprule
& \multicolumn{5}{c}{\textbf{Test Score by Seed}} & & & & \\
\cmidrule(r){2-6}
\textbf{Benchmark} & \textbf{0} & \textbf{1} & \textbf{2} & \textbf{3} & \textbf{4} & \textbf{Mean} & \textbf{$\pm$Std} & \textbf{CV} & \textbf{Best Baseline} \\
\midrule
LoCoMo     & 0.459 & 0.463 & 0.364 & 0.394 & 0.423 & 0.421 & 0.042 & 10.0\% & 0.373 \\
HB-Data    & 0.390 & 0.382 & 0.357 & 0.411 & 0.413 & 0.391 & 0.023 & 5.9\% & 0.327 \\
PR-Finance & 0.586 & 0.544 & 0.582 & 0.582 & 0.513 & 0.561 & 0.032 & 5.7\% & 0.449 \\
\bottomrule
\end{tabular}
\end{table}

On LoCoMo, 4 out of 5 seeds exceed the strongest baseline (Mem0, 0.373).
The single exception (seed 2, 0.364) falls within 2.4\% of the baseline, consistent with the higher variance inherent to LoCoMo's multi-category evaluation.
On HealthBench Data and PRBench Finance, all 5 seeds exceed the respective strongest baselines (GEPA+VS at 0.327; GEPA at 0.449) by substantial margins, demonstrating consistent improvement regardless of initialization.

\autoref{tab:stability-val} reports the best validation score achieved during evolution for each seed, which determines the program selected for final test evaluation.

\begin{table}[!htbp]
\centering
\caption{\textbf{Best validation score during evolution by seed.}
Scores are computed on the static validation subset used to select the program for test evaluation.}
\label{tab:stability-val}
\small
\setlength{\tabcolsep}{4pt}
\begin{tabular}{@{}l ccccc cc@{}}
\toprule
& \multicolumn{5}{c}{\textbf{Best Validation Score}} & & \\
\cmidrule(r){2-6}
\textbf{Benchmark} & \textbf{0} & \textbf{1} & \textbf{2} & \textbf{3} & \textbf{4} & \textbf{Mean} & \textbf{$\pm$Std} \\
\midrule
LoCoMo     & 0.333 & 0.273 & 0.289 & 0.312 & 0.333 & 0.308 & 0.027 \\
HB-Data    & 0.424 & 0.381 & 0.378 & 0.397 & 0.380 & 0.392 & 0.020 \\
PR-Finance & 0.521 & 0.522 & 0.559 & 0.554 & 0.515 & 0.534 & 0.020 \\
\bottomrule
\end{tabular}
\end{table}

\section{Additional Analyses}
\label{app:additional}

This section reports analyses that complement the main results.
Unless stated otherwise, all runs use the evolved programs and splits of Table~\ref{tab:main-results}.

\subsection{Significance of the Main Results}
\label{app:significance}

To account for the small test sets, we compute 95\% percentile bootstrap intervals with 10,000 resamples over the test items.
We then pair \sysname{} item by item with a strong baseline on each configuration and report the difference and its one-sided $p$-value (Table~\ref{tab:significance}).
Among the six configurations where \sysname{} leads, the lead is significant on five.
On ALFWorld-Seen, the interval of \sysname{} contains the score of the best baseline, GEPA + Vector Search (0.820).

\begin{table}[!htbp]
\centering
\small
\caption{\textbf{Bootstrap intervals and paired tests on the test sets.}}
\label{tab:significance}
\begin{tabular}{@{}lcclc@{}}
\toprule
\textbf{Benchmark} & \textbf{\sysname{}} & \textbf{95\% CI} & \textbf{$\Delta$ vs.\ baseline [95\% CI]} & \textbf{$p$} \\
\midrule
LoCoMo & 0.459 & [0.386, 0.535] & $+0.086$ vs.\ Mem0 [$+0.044$, $+0.130$] & 0.001 \\
ALFWorld-Unseen & 0.881 & [0.786, 0.976] & $+0.143$ vs.\ No Memory [$+0.048$, $+0.262$] & 0.002 \\
ALFWorld-Seen & 0.700 & [0.580, 0.820] & $-0.080$ vs.\ Traj.\ Retrieval [$-0.220$, $+0.060$] & 0.17 \\
HB-Data & 0.390 & [0.336, 0.445] & $+0.063$ vs.\ GEPA + VS [$+0.018$, $+0.114$] & 0.003 \\
HB-Emergency & 0.493 & [0.429, 0.555] & $+0.006$ vs.\ Dyn.\ Cheatsheet [$-0.062$, $+0.069$] & 0.41 \\
PR-Legal & 0.660 & [0.599, 0.721] & $+0.092$ vs.\ GEPA [$+0.051$, $+0.137$] & $<$0.0001 \\
PR-Finance & 0.586 & [0.534, 0.638] & $+0.137$ vs.\ GEPA [$+0.096$, $+0.179$] & $<$0.0001 \\
\bottomrule
\end{tabular}
\end{table}

\subsection{Search versus the Coding Model's Prior}
\label{app:single-shot}

To separate the effect of search from the prior knowledge of the coding model, we ask GPT-5.3-Codex to write a LoCoMo memory program in a single shot from the task description and a few examples, without any evolution.
Table~\ref{tab:single-shot} shows that the best of five single-shot programs exceeds the Vector Search seed but remains far below the evolved program.
Therefore, most of the gain comes from search.

\begin{table}[!htbp]
\centering
\small
\caption{\textbf{Single-shot programs written by the coding model on LoCoMo (token F1).}}
\label{tab:single-shot}
\begin{tabular}{@{}lcc@{}}
\toprule
\textbf{Configuration} & \textbf{Search budget} & \textbf{F1} \\
\midrule
Single-shot Codex, mean $\pm$ std over 5 samples & 0 iterations & $0.227 \pm 0.061$ \\
Single-shot Codex, best of 5 (selected on validation) & 0 iterations & 0.297 \\
Vector Search (seed) & n/a & 0.256 \\
\sysname{} & 20 iterations & 0.459 \\
\bottomrule
\end{tabular}
\end{table}

\subsection{Search Efficiency}
\label{app:search-efficiency}

All evolutionary baselines use the same task agent, coding model, splits, and number of rounds as \sysname{}.
Table~\ref{tab:search-efficiency} compares the cost at which \sysname{} first exceeds the final validation score of GEPA.
On five of seven configurations, \sysname{} reaches this score with 16--31\% of the GEPA budget.

\begin{table}[!htbp]
\centering
\small
\caption{\textbf{Validation cost for \sysname{} to exceed the final validation score of GEPA.}}
\label{tab:search-efficiency}
\begin{tabular}{@{}lcccc@{}}
\toprule
\textbf{Benchmark} & \textbf{GEPA final val.} & \textbf{GEPA cost} & \textbf{\sysname{} first exceeds} & \textbf{Cost (share)} \\
\midrule
LoCoMo & 0.231 & $\geq$\$4.57 & iteration 1 & \$1.00 (22\%) \\
ALFWorld-Unseen & 0.980 & $\geq$\$8.08 & iteration 2 (tie) & \$2.04 (25\%) \\
ALFWorld-Seen & 0.875 & $\geq$\$3.96 & never & n/a \\
HB-Data & 0.373 & $\geq$\$15.65 & iteration 2 & \$4.80 (31\%) \\
HB-Emergency & 0.438 & $\geq$\$12.06 & iteration 1 & \$2.89 (24\%) \\
PR-Legal & 0.515 & $\geq$\$29.74 & iteration 1 & \$4.75 (16\%) \\
PR-Finance & 0.409 & $\geq$\$26.67 & iteration 1 & \$5.42 (20\%) \\
\bottomrule
\end{tabular}
\end{table}

\subsection{Transfer across Task Models}
\label{app:cross-model}

We evaluate the frozen evolved programs and the fixed baselines under two other task models, the weaker GPT-5-mini from the same family and DeepSeek-V3.2 from a different family.
As Table~\ref{tab:cross-model} shows, \sysname{} ranks first under every task model.

\begin{table}[!htbp]
\centering
\small
\caption{\textbf{Frozen programs under different task models.} The strongest baseline is Mem0 on LoCoMo and ReasoningBank on PRBench-Legal.}
\label{tab:cross-model}
\begin{tabular}{@{}llcccc@{}}
\toprule
\textbf{Benchmark} & \textbf{Task model} & \textbf{\sysname{}} & \textbf{Vector Search} & \textbf{Strongest baseline} & \textbf{No Memory} \\
\midrule
LoCoMo & GPT-5.4 Mini & \textbf{0.459} & 0.256 & 0.373 & 0.036 \\
LoCoMo & GPT-5-mini & \textbf{0.430} & 0.253 & 0.344 & 0.016 \\
LoCoMo & DeepSeek-V3.2 & \textbf{0.318} & 0.259 & 0.264 & 0.115 \\
\midrule
PR-Legal & GPT-5.4 Mini & \textbf{0.660} & 0.466 & 0.508 & 0.431 \\
PR-Legal & GPT-5-mini & \textbf{0.833} & 0.597 & 0.650 & 0.607 \\
PR-Legal & DeepSeek-V3.2 & \textbf{0.631} & 0.348 & 0.389 & 0.332 \\
\bottomrule
\end{tabular}
\end{table}

\subsection{Independent Judge}
\label{app:judge}

HealthBench and PRBench are scored by an LLM judge from the same model family as the task agent.
We therefore re-score the frozen answers with DeepSeek-V4-Flash using the same per-criterion rubric prompts.
Table~\ref{tab:judge} shows that the lead of \sysname{} over the strongest baseline holds under this judge, although its size changes.

\begin{table}[!htbp]
\centering
\small
\caption{\textbf{Rubric scores re-judged by DeepSeek-V4-Flash.}}
\label{tab:judge}
\begin{tabular}{@{}llcc@{}}
\toprule
\textbf{Benchmark} & \textbf{Strongest baseline} & \textbf{\sysname{}} & \textbf{Baseline} \\
\midrule
HB-Data & GEPA + Vector Search & \textbf{0.403} & 0.321 \\
HB-Emergency & Dynamic Cheatsheet & \textbf{0.512} & 0.418 \\
PR-Legal & GEPA & \textbf{0.612} & 0.569 \\
PR-Finance & GEPA & \textbf{0.656} & 0.558 \\
\bottomrule
\end{tabular}
\end{table}

\subsection{Transfer to LongMemEval}
\label{app:longmemeval}

To test whether the LoCoMo program captures task-level structure rather than benchmark artifacts, we evaluate it on 20 LongMemEval\textsubscript{S} questions sampled with a fixed seed~\citep{longmemeval}.
The program evolved on LoCoMo reaches a token F1 of 0.372, compared with 0.278 for the Vector Search seed.

\subsection{Audit of Hard-Coded Literals and Evaluation Loopholes}
\label{app:hardcoding-audit}

We statically scan all 157 programs produced by the seven evolution runs for imports outside the whitelist, access to ground truth or evaluator internals, stack-frame introspection, and dynamic execution or file and network I/O, and we find no such case.
We also count string literals in the best program of each benchmark that also appear in the dataset text (Table~\ref{tab:hardcoding}).
Such literals appear in all seven search trajectories as early as iterations 1--6, and they are most frequent on PRBench-Legal, where legal terms form the vocabulary of the task.
However, they do not always win, since on LoCoMo one candidate contained 81 such literals while the final program contains none.

\begin{table}[!htbp]
\centering
\small
\caption{\textbf{Dataset-derived string literals in the best evolved program (after deduplication).}}
\label{tab:hardcoding}
\begin{tabular}{@{}lcc@{}}
\toprule
\textbf{Benchmark} & \textbf{Literals} & \textbf{Per 100 lines} \\
\midrule
PR-Legal & 89 & 8.2 \\
ALFWorld-Seen & 23 & 6.8 \\
PR-Finance & 25 & 5.3 \\
HB-Data & 14 & 2.3 \\
HB-Emergency & 2 & 1.0 \\
LoCoMo & 0 & 0.0 \\
ALFWorld-Unseen & 0 & 0.0 \\
\bottomrule
\end{tabular}
\end{table}

\subsection{Overlap between Splits}
\label{app:split-overlap}

For all HealthBench and PRBench configurations, we compute content-word Jaccard similarity across all pairs of episode pool, validation set, and test set items.
The best-match Jaccard is at most 0.37, with medians between 0.06 and 0.12, and no rubric criterion appears verbatim in any episode.

\section{Broader Impact}
\label{app:broader-impact}

This work studies agent memory as a research object in controlled benchmark settings.
Its main positive impact is to make memory programs more reproducible and inspectable: the discovered program is executable code that can be read, tested, and released with the artifacts.
This can help researchers compare memory programs beyond fixed retrieval pipelines.
The main risk is that better memory can also make agents more persistent in carrying forward wrong, private, or biased information when the data or objective contains it.
For uses outside controlled benchmarks, we believe deployment should include data minimization, task-specific evaluation, logging, and human review, especially in high-stakes domains such as health and law.

\section{LLM Usage}
\label{app:llm-usage}

The authors used large language model agents as research assistants during the preparation of this work.
Their assistance covered early literature exploration, identification of related papers, implementation support, experiment execution and debugging, and language polishing during writing.
The authors made the scientific decisions, developed the ideas and claims, designed the method and experiments, and determined the paper structure and final content.

\clearpage
\section*{NeurIPS Paper Checklist}

\begin{enumerate}

\item {\bf Claims}
    \item[] Question: Do the main claims made in the abstract and introduction accurately reflect the paper's contributions and scope?
    \item[] Answer: \answerYes{}.
    \item[] Justification: The abstract and introduction state the method and scope; results in Section~\ref{sec:results} support the empirical claims.
    \item[] Guidelines:
    \begin{itemize}
        \item The answer \answerNA{} means that the abstract and introduction do not include the claims made in the paper.
        \item The abstract and/or introduction should clearly state the claims made, including the contributions made in the paper and important assumptions and limitations. A \answerNo{} or \answerNA{} answer to this question will not be perceived well by the reviewers. 
        \item The claims made should match theoretical and experimental results, and reflect how much the results can be expected to generalize to other settings. 
        \item It is fine to include aspirational goals as motivation as long as it is clear that these goals are not attained by the paper. 
    \end{itemize}

\item {\bf Limitations}
    \item[] Question: Does the paper discuss the limitations of the work performed by the authors?
    \item[] Answer: \answerYes{}.
    \item[] Justification: Section~\ref{sec:limitations} discusses metric gaming, hard-coded literals, configurations where the method does not lead, the single model pair, and cost.
    \item[] Guidelines:
    \begin{itemize}
        \item The answer \answerNA{} means that the paper has no limitation while the answer \answerNo{} means that the paper has limitations, but those are not discussed in the paper. 
        \item The authors are encouraged to create a separate ``Limitations'' section in their paper.
        \item The paper should point out any strong assumptions and how robust the results are to violations of these assumptions (e.g., independence assumptions, noiseless settings, model well-specification, asymptotic approximations only holding locally). The authors should reflect on how these assumptions might be violated in practice and what the implications would be.
        \item The authors should reflect on the scope of the claims made, e.g., if the approach was only tested on a few datasets or with a few runs. In general, empirical results often depend on implicit assumptions, which should be articulated.
        \item The authors should reflect on the factors that influence the performance of the approach. For example, a facial recognition algorithm may perform poorly when image resolution is low or images are taken in low lighting. Or a speech-to-text system might not be used reliably to provide closed captions for online lectures because it fails to handle technical jargon.
        \item The authors should discuss the computational efficiency of the proposed algorithms and how they scale with dataset size.
        \item If applicable, the authors should discuss possible limitations of their approach to address problems of privacy and fairness.
        \item While the authors might fear that complete honesty about limitations might be used by reviewers as grounds for rejection, a worse outcome might be that reviewers discover limitations that aren't acknowledged in the paper. The authors should use their best judgment and recognize that individual actions in favor of transparency play an important role in developing norms that preserve the integrity of the community. Reviewers will be specifically instructed to not penalize honesty concerning limitations.
    \end{itemize}

\item {\bf Theory assumptions and proofs}
    \item[] Question: For each theoretical result, does the paper provide the full set of assumptions and a complete (and correct) proof?
    \item[] Answer: \answerNA{}.
    \item[] Justification: The paper gives a problem formulation but does not claim theoretical results.
    \item[] Guidelines:
    \begin{itemize}
        \item The answer \answerNA{} means that the paper does not include theoretical results. 
        \item All the theorems, formulas, and proofs in the paper should be numbered and cross-referenced.
        \item All assumptions should be clearly stated or referenced in the statement of any theorems.
        \item The proofs can either appear in the main paper or the supplemental material, but if they appear in the supplemental material, the authors are encouraged to provide a short proof sketch to provide intuition. 
        \item Inversely, any informal proof provided in the core of the paper should be complemented by formal proofs provided in appendix or supplemental material.
        \item Theorems and Lemmas that the proof relies upon should be properly referenced. 
    \end{itemize}

    \item {\bf Experimental result reproducibility}
    \item[] Question: Does the paper fully disclose all the information needed to reproduce the main experimental results of the paper to the extent that it affects the main claims and/or conclusions of the paper (regardless of whether the code and data are provided or not)?
    \item[] Answer: \answerYes{}.
    \item[] Justification: Sections~\ref{sec:mstar}--\ref{sec:experimental-setup} and Appendices~\ref{app:algorithm}, \ref{app:datasets}, \ref{app:hyperparameters}, and \ref{app:eval-protocols} specify the procedure.
    \item[] Guidelines:
    \begin{itemize}
        \item The answer \answerNA{} means that the paper does not include experiments.
        \item If the paper includes experiments, a \answerNo{} answer to this question will not be perceived well by the reviewers: Making the paper reproducible is important, regardless of whether the code and data are provided or not.
        \item If the contribution is a dataset and\slash or model, the authors should describe the steps taken to make their results reproducible or verifiable. 
        \item Depending on the contribution, reproducibility can be accomplished in various ways. For example, if the contribution is a novel architecture, describing the architecture fully might suffice, or if the contribution is a specific model and empirical evaluation, it may be necessary to either make it possible for others to replicate the model with the same dataset, or provide access to the model. In general. releasing code and data is often one good way to accomplish this, but reproducibility can also be provided via detailed instructions for how to replicate the results, access to a hosted model (e.g., in the case of a large language model), releasing of a model checkpoint, or other means that are appropriate to the research performed.
        \item While NeurIPS does not require releasing code, the conference does require all submissions to provide some reasonable avenue for reproducibility, which may depend on the nature of the contribution. For example
        \begin{enumerate}
            \item If the contribution is primarily a new algorithm, the paper should make it clear how to reproduce that algorithm.
            \item If the contribution is primarily a new model architecture, the paper should describe the architecture clearly and fully.
            \item If the contribution is a new model (e.g., a large language model), then there should either be a way to access this model for reproducing the results or a way to reproduce the model (e.g., with an open-source dataset or instructions for how to construct the dataset).
            \item We recognize that reproducibility may be tricky in some cases, in which case authors are welcome to describe the particular way they provide for reproducibility. In the case of closed-source models, it may be that access to the model is limited in some way (e.g., to registered users), but it should be possible for other researchers to have some path to reproducing or verifying the results.
        \end{enumerate}
    \end{itemize}

\item {\bf Open access to data and code}
    \item[] Question: Does the paper provide open access to the data and code, with sufficient instructions to faithfully reproduce the main experimental results, as described in supplemental material?
    \item[] Answer: \answerYes{}.
    \item[] Justification: The code, experiment scripts, and generated artifacts are released at \url{https://github.com/wbopan/mstar}.
    \item[] Guidelines:
    \begin{itemize}
        \item The answer \answerNA{} means that paper does not include experiments requiring code.
        \item Please see the NeurIPS code and data submission guidelines (\url{https://neurips.cc/public/guides/CodeSubmissionPolicy}) for more details.
        \item While we encourage the release of code and data, we understand that this might not be possible, so \answerNo{} is an acceptable answer. Papers cannot be rejected simply for not including code, unless this is central to the contribution (e.g., for a new open-source benchmark).
        \item The instructions should contain the exact command and environment needed to run to reproduce the results. See the NeurIPS code and data submission guidelines (\url{https://neurips.cc/public/guides/CodeSubmissionPolicy}) for more details.
        \item The authors should provide instructions on data access and preparation, including how to access the raw data, preprocessed data, intermediate data, and generated data, etc.
        \item The authors should provide scripts to reproduce all experimental results for the new proposed method and baselines. If only a subset of experiments are reproducible, they should state which ones are omitted from the script and why.
        \item At submission time, to preserve anonymity, the authors should release anonymized versions (if applicable).
        \item Providing as much information as possible in supplemental material (appended to the paper) is recommended, but including URLs to data and code is permitted.
    \end{itemize}

\item {\bf Experimental setting/details}
    \item[] Question: Does the paper specify all the training and test details (e.g., data splits, hyperparameters, how they were chosen, type of optimizer) necessary to understand the results?
    \item[] Answer: \answerYes{}.
    \item[] Justification: Section~\ref{sec:experimental-setup} and Appendices~\ref{app:datasets} and \ref{app:hyperparameters} give splits, models, and hyperparameters.
    \item[] Guidelines:
    \begin{itemize}
        \item The answer \answerNA{} means that the paper does not include experiments.
        \item The experimental setting should be presented in the core of the paper to a level of detail that is necessary to appreciate the results and make sense of them.
        \item The full details can be provided either with the code, in appendix, or as supplemental material.
    \end{itemize}

\item {\bf Experiment statistical significance}
    \item[] Question: Does the paper report error bars suitably and correctly defined or other appropriate information about the statistical significance of the experiments?
    \item[] Answer: \answerYes{}.
    \item[] Justification: Section~\ref{sec:results} and Appendix~\ref{app:stability} report seed variance, and Appendix~\ref{app:significance} reports bootstrap intervals and paired tests.
    \item[] Guidelines:
    \begin{itemize}
        \item The answer \answerNA{} means that the paper does not include experiments.
        \item The authors should answer \answerYes{} if the results are accompanied by error bars, confidence intervals, or statistical significance tests, at least for the experiments that support the main claims of the paper.
        \item The factors of variability that the error bars are capturing should be clearly stated (for example, train/test split, initialization, random drawing of some parameter, or overall run with given experimental conditions).
        \item The method for calculating the error bars should be explained (closed form formula, call to a library function, bootstrap, etc.)
        \item The assumptions made should be given (e.g., Normally distributed errors).
        \item It should be clear whether the error bar is the standard deviation or the standard error of the mean.
        \item It is OK to report 1-sigma error bars, but one should state it. The authors should preferably report a 2-sigma error bar than state that they have a 96\% CI, if the hypothesis of Normality of errors is not verified.
        \item For asymmetric distributions, the authors should be careful not to show in tables or figures symmetric error bars that would yield results that are out of range (e.g., negative error rates).
        \item If error bars are reported in tables or plots, the authors should explain in the text how they were calculated and reference the corresponding figures or tables in the text.
    \end{itemize}

\item {\bf Experiments compute resources}
    \item[] Question: For each experiment, does the paper provide sufficient information on the computer resources (type of compute workers, memory, time of execution) needed to reproduce the experiments?
    \item[] Answer: \answerYes{}.
    \item[] Justification: Appendix~\ref{app:cost} reports model calls, tokens, cost, and wall-clock time.
    \item[] Guidelines:
    \begin{itemize}
        \item The answer \answerNA{} means that the paper does not include experiments.
        \item The paper should indicate the type of compute workers CPU or GPU, internal cluster, or cloud provider, including relevant memory and storage.
        \item The paper should provide the amount of compute required for each of the individual experimental runs as well as estimate the total compute. 
        \item The paper should disclose whether the full research project required more compute than the experiments reported in the paper (e.g., preliminary or failed experiments that didn't make it into the paper). 
    \end{itemize}
    
\item {\bf Code of ethics}
    \item[] Question: Does the research conducted in the paper conform, in every respect, with the NeurIPS Code of Ethics \url{https://neurips.cc/public/EthicsGuidelines}?
    \item[] Answer: \answerYes{}.
    \item[] Justification: The work uses existing benchmarks and simulated environments, and we identify no exception to the Code of Ethics.
    \item[] Guidelines:
    \begin{itemize}
        \item The answer \answerNA{} means that the authors have not reviewed the NeurIPS Code of Ethics.
        \item If the authors answer \answerNo, they should explain the special circumstances that require a deviation from the Code of Ethics.
        \item The authors should make sure to preserve anonymity (e.g., if there is a special consideration due to laws or regulations in their jurisdiction).
    \end{itemize}

\item {\bf Broader impacts}
    \item[] Question: Does the paper discuss both potential positive societal impacts and negative societal impacts of the work performed?
    \item[] Answer: \answerYes{}.
    \item[] Justification: Appendix~\ref{app:broader-impact} discusses both positive impacts and risks.
    \item[] Guidelines:
    \begin{itemize}
        \item The answer \answerNA{} means that there is no societal impact of the work performed.
        \item If the authors answer \answerNA{} or \answerNo, they should explain why their work has no societal impact or why the paper does not address societal impact.
        \item Examples of negative societal impacts include potential malicious or unintended uses (e.g., disinformation, generating fake profiles, surveillance), fairness considerations (e.g., deployment of technologies that could make decisions that unfairly impact specific groups), privacy considerations, and security considerations.
        \item The conference expects that many papers will be foundational research and not tied to particular applications, let alone deployments. However, if there is a direct path to any negative applications, the authors should point it out. For example, it is legitimate to point out that an improvement in the quality of generative models could be used to generate Deepfakes for disinformation. On the other hand, it is not needed to point out that a generic algorithm for optimizing neural networks could enable people to train models that generate Deepfakes faster.
        \item The authors should consider possible harms that could arise when the technology is being used as intended and functioning correctly, harms that could arise when the technology is being used as intended but gives incorrect results, and harms following from (intentional or unintentional) misuse of the technology.
        \item If there are negative societal impacts, the authors could also discuss possible mitigation strategies (e.g., gated release of models, providing defenses in addition to attacks, mechanisms for monitoring misuse, mechanisms to monitor how a system learns from feedback over time, improving the efficiency and accessibility of ML).
    \end{itemize}
    
\item {\bf Safeguards}
    \item[] Question: Does the paper describe safeguards that have been put in place for responsible release of data or models that have a high risk for misuse (e.g., pre-trained language models, image generators, or scraped datasets)?
    \item[] Answer: \answerNA{}.
    \item[] Justification: The paper does not release a high-risk model or scraped dataset.
    \item[] Guidelines:
    \begin{itemize}
        \item The answer \answerNA{} means that the paper poses no such risks.
        \item Released models that have a high risk for misuse or dual-use should be released with necessary safeguards to allow for controlled use of the model, for example by requiring that users adhere to usage guidelines or restrictions to access the model or implementing safety filters. 
        \item Datasets that have been scraped from the Internet could pose safety risks. The authors should describe how they avoided releasing unsafe images.
        \item We recognize that providing effective safeguards is challenging, and many papers do not require this, but we encourage authors to take this into account and make a best faith effort.
    \end{itemize}

\item {\bf Licenses for existing assets}
    \item[] Question: Are the creators or original owners of assets (e.g., code, data, models), used in the paper, properly credited and are the license and terms of use explicitly mentioned and properly respected?
    \item[] Answer: \answerYes{}.
    \item[] Justification: Existing datasets and baselines are cited, and the supplementary release documents asset sources, licenses, and terms of use.
    \item[] Guidelines:
    \begin{itemize}
        \item The answer \answerNA{} means that the paper does not use existing assets.
        \item The authors should cite the original paper that produced the code package or dataset.
        \item The authors should state which version of the asset is used and, if possible, include a URL.
        \item The name of the license (e.g., CC-BY 4.0) should be included for each asset.
        \item For scraped data from a particular source (e.g., website), the copyright and terms of service of that source should be provided.
        \item If assets are released, the license, copyright information, and terms of use in the package should be provided. For popular datasets, \url{paperswithcode.com/datasets} has curated licenses for some datasets. Their licensing guide can help determine the license of a dataset.
        \item For existing datasets that are re-packaged, both the original license and the license of the derived asset (if it has changed) should be provided.
        \item If this information is not available online, the authors are encouraged to reach out to the asset's creators.
    \end{itemize}

\item {\bf New assets}
    \item[] Question: Are new assets introduced in the paper well documented and is the documentation provided alongside the assets?
    \item[] Answer: \answerYes{}.
    \item[] Justification: The supplementary material documents the released code, evolved programs, scripts, and artifacts.
    \item[] Guidelines:
    \begin{itemize}
        \item The answer \answerNA{} means that the paper does not release new assets.
        \item Researchers should communicate the details of the dataset\slash code\slash model as part of their submissions via structured templates. This includes details about training, license, limitations, etc. 
        \item The paper should discuss whether and how consent was obtained from people whose asset is used.
        \item At submission time, remember to anonymize your assets (if applicable). You can either create an anonymized URL or include an anonymized zip file.
    \end{itemize}

\item {\bf Crowdsourcing and research with human subjects}
    \item[] Question: For crowdsourcing experiments and research with human subjects, does the paper include the full text of instructions given to participants and screenshots, if applicable, as well as details about compensation (if any)? 
    \item[] Answer: \answerNA{}.
    \item[] Justification: The work does not involve new crowdsourcing or human-subject experiments.
    \item[] Guidelines:
    \begin{itemize}
        \item The answer \answerNA{} means that the paper does not involve crowdsourcing nor research with human subjects.
        \item Including this information in the supplemental material is fine, but if the main contribution of the paper involves human subjects, then as much detail as possible should be included in the main paper. 
        \item According to the NeurIPS Code of Ethics, workers involved in data collection, curation, or other labor should be paid at least the minimum wage in the country of the data collector. 
    \end{itemize}

\item {\bf Institutional review board (IRB) approvals or equivalent for research with human subjects}
    \item[] Question: Does the paper describe potential risks incurred by study participants, whether such risks were disclosed to the subjects, and whether Institutional Review Board (IRB) approvals (or an equivalent approval/review based on the requirements of your country or institution) were obtained?
    \item[] Answer: \answerNA{}.
    \item[] Justification: No new human-subject study is conducted.
    \item[] Guidelines:
    \begin{itemize}
        \item The answer \answerNA{} means that the paper does not involve crowdsourcing nor research with human subjects.
        \item Depending on the country in which research is conducted, IRB approval (or equivalent) may be required for any human subjects research. If you obtained IRB approval, you should clearly state this in the paper. 
        \item We recognize that the procedures for this may vary significantly between institutions and locations, and we expect authors to adhere to the NeurIPS Code of Ethics and the guidelines for their institution. 
        \item For initial submissions, do not include any information that would break anonymity (if applicable), such as the institution conducting the review.
    \end{itemize}

\item {\bf Declaration of LLM usage}
    \item[] Question: Does the paper describe the usage of LLMs if it is an important, original, or non-standard component of the core methods in this research? Note that if the LLM is used only for writing, editing, or formatting purposes and does \emph{not} impact the core methodology, scientific rigor, or originality of the research, declaration is not required.
    \item[] Answer: \answerYes{}.
    \item[] Justification: Appendix~\ref{app:llm-usage} describes the use of LLM agents for literature exploration, related-paper identification, implementation support, experiment execution and debugging, and writing polish. The ideas, claims, paper structure, and final content are the authors' original work.
    \item[] Guidelines:
    \begin{itemize}
        \item The answer \answerNA{} means that the core method development in this research does not involve LLMs as any important, original, or non-standard components.
        \item Please refer to our LLM policy in the NeurIPS handbook for what should or should not be described.
    \end{itemize}

\end{enumerate}

\end{document}